\documentclass[12pt]{iopart}
\usepackage{iopams}
\usepackage{graphicx}
\usepackage{amssymb}
  
\begin{document}

\title[
First hitting times of RWs on directed ER networks
]{
The distribution of first hitting times
of random walks on directed Erd\H{o}s-R\'enyi networks
}

\author{Ido Tishby, Ofer Biham \& Eytan Katzav}
\address{Racah Institute of Physics, 
The Hebrew University, Jerusalem 91904, Israel.}
\eads{\mailto{ido.tishby@mail.huji.ac.il}, \mailto{biham@phys.huji.ac.il}, 
\mailto{eytan.katzav@mail.huji.ac.il}}

\begin{abstract}
We present analytical results for   
the distribution of first hitting times 
of random walkers (RWs) on 
directed 
Erd\H{o}s-R\'enyi (ER) networks.
Starting from a random initial node,
a random walker hops randomly 
along directed edges
between adjacent nodes in the network.
The path terminates 
either by the retracing scenario,
when the walker
enters a node which 
it has already visited before,
or by the trapping scenario, when it
becomes trapped in a dead-end node 
from which it cannot exit.
The path length, namely
the number of steps, $d$, pursued 
by the random walker from the initial node
up to its termination, is called 
the first hitting time.
Using recursion equations, 
we obtain 
analytical results for the tail distribution of first hitting times, 
$P(d > \ell)$.
The results are found to be in excellent agreement with numerical
simulations.
It turns out that the distribution
$P(d > \ell)$
can be expressed as a product of an exponential distribution 
and a Rayleigh distribution.
We obtain expressions for the mean, median and standard deviation 
of this distribution in terms of the network size and its mean degree.
We also calculate the distribution of last hitting times,
namely the path lengths of self-avoiding walks
on directed ER networks, which do not
retrace their paths.
The last hitting times are found to be much longer
than the first hitting times.
The results are compared to those obtained for undirected 
ER networks.
It is found that the first hitting times of RWs in a directed
ER network are much longer than in the 
corresponding undirected network.
This is due to the fact that
RWs on directed networks do not exhibit the backtracking
scenario, which is a dominant termination
mechanism of RWs on undirected networks.
It is shown that our approach also applies to a broader 
class of networks,
referred to as semi-ER networks, in which
the distribution of in-degrees is Poisson,
while the out-degrees may follow any desired distribution
with the same mean as the in-degree distribution.
\end{abstract}

\section{Introduction}

Random walk (RW) models
provide a framework for the study of diffusion
and other stochastic processes
\cite{Lawler2010,Rudnick2010}.
These models were studied extensively on 
regular lattices and random networks.
An RW can be considered as a particle which resides
on the sites of the lattice or network, such that at
each time step it hops randomly to one of the
neighbors of its current site.
Random walks on lattices provide a discrete spatio-temporal description of diffusion
processes in the Euclidean space.  
In the large-scale and long-time limit, 
the evolution of the spatial probability distribution of an RW 
can be described by the diffusion equation, while
at small scales, the discreteness of the lattice plays an important role.
Unlike a ballistic particle which moves in a fixed direction at a constant speed,
an RW picks a random direction at each time step.
As a result, the mean distance reached by an RW from its initial location
scales like $t^{1/2}$, compared to $t$ for a ballistic particle, where $t$ is
the elapsed time.
More precisely, the probability distribution of an RW starting at the origin
of a regular lattice in $D$ dimensions follows a Gaussian distribution,
whose standard deviation scales like $t^{1/2}$,
for any finite dimension, $D$.
RWs maintain no memory in the sense that the probability distribution
of the next move depends only on the current state of the system.
Therefore, they satisfy the Markovian condition and can be
studied using the methodologies developed for Markovian processes
\cite{Kampen2007}.
RW problems are commonly posed as initial value problems,
in which the initial conditions are specified and the task is to
calculate the spatial probability distribution at a later time.
However, in an alternative setting called first passage problems,
one is interested in questions like how long it will take for the
RW to reach a given location for the first time
\cite{Redner2001}. 
The general
problem is to calculate the distribution of first passage times
for the given setting, or properties of the distribution such 
as the mean first passage time.
An RW on a lattice hops randomly at each time step
to one of the nearest neighbors of its current site.
In some of the steps it hops into new sites which 
have not been visited before.
In other steps it hops into previously visited sites.
The mean number of distinct sites, $s(t)$, visited up to 
time $t$ is thus smaller than $t$. It was shown that in
one dimension $s(t) \sim t^{1/2}$, 
in two dimensions $s(t) \sim t/\ln t$, while in three
and more dimensions $s(t) \sim t$
\cite{Montroll1965}.
It was recently shown that on ER networks,
the number of distinct nodes visited up to time $t$
scales linearly with $t$
\cite{Debacco2015},
resembling the results obtained
for RWs on high dimensional lattices.
On finite networks other interesting quantities emerge, 
such as the mean first passage time between a random pair of nodes
\cite{Sood2005} and
the mean cover time, namely the average number of steps required
for the RW to visit all the nodes in the network
\cite{Kahn1989}.

An important time scale which appears 
in random walks on networks is the
{\it first hitting time},
also referred to as the first intersection length
\cite{Herrero2003,Herrero2005,Herrero2005b,Herrero2007}.
This time scale emerges in a class of RW models 
in which the RW keeps hopping
until it enters a node which it has already visited
before or becomes trapped in a node from which it cannot exit. 
At this point the path is terminated,
as is the case in a large class of processes called
first passage problems
\cite{Redner2001}.
The resulting path length, namely 
the number of time steps up to its termination,
is called the first hitting time.
In Ref.
\cite{Tishby2016b}
we presented analytical results for the
distribution 
of first hitting times of RWs on undirected
Erd{\H o}s-R\'enyi 
(ER) networks
\cite{Erdos1959,Erdos1960,Erdos1961}.
On undirected networks, 
the RW path may terminate either by 
backtracking into the previous node
or by retracing itself, namely stepping into a node which was 
already visited two or more time steps earlier.
By calculating the probabilities of these two scenarios, we
obtained analytical results for the distribution of
first hitting times of RWs on ER networks.
Another interesting time scale is the 
{\it last hitting time}
\cite{Herrero2005}
of a self avoiding walk (SAW)
\cite{Madras1996,Slade2011}, 
which does not retrace its path
but may be trapped once it enters a node which does not have
any yet unvisited neighbors.
In Ref.
\cite{Tishby2016a}
we presented analytical results for the distribution of last hitting times
of SAWs on undirected ER networks.

Most of the networks encountered in physical, chemical, biological,
technological and social systems are directed networks.
Therefore, it is important to study diffusive processes on directed networks
and the RW is the simplest dynamical model describing such processes.
The dynamical properties of RWs on directed networks are different from
those of RWs on undirected networks.
In undirected networks each RW path between node $i$ and node $j$
may be pursued in both directions. 
In directed networks some pairs of nodes may be connected only
in one direction and not in the other direction. 
Even if nodes $i$ and $j$ are connected in both directions, 
the paths in opposite directions are not the same.
While undirected networks exhibit a single degree distribution,
$p(k)$, directed networks exhibit two distinct degree distributions, 
namely the distribution of in-degrees, $p^{\rm in}(k)$, and 
the distribution of out-degrees, $p^{\rm out}(k)$.
The two distributions must have the same mean, namely
$\langle k \rangle^{\rm in}=\langle k \rangle^{\rm out}$.
While in undirected networks the frequency in which an RW
visits a node, $i$, is simply proportional to its degree, $k_i$,
in directed networks such visit frequencies depend not only
on the in-degree $k_i^{\rm in}$ but on the overall structure
of the surrounding network.
Finally, directed networks exhibit dead-end nodes which have incoming
links but no outgoing links. As a result, RWs which enter such
nodes becomes trapped. 

In this paper we present analytical results for 
the distribution 
of first hitting times of RWs
on {\it directed} ER networks
\cite{Bollobas2001,Bang-Jensen2007,Graham2008}.
In these networks each pair of nodes, $i$ and $j$,
are connected, with probability $p$ by a directed link from
$i$ to $j$, and independently, with the same probability,
by a directed link from $j$ to $i$.
We also calculate the mean, median and
standard deviation
of the distribution of first hitting times.
The results are found to be in excellent agreement with numerical
simulations.
Unlike the case of undirected networks in which backtracking moves
are always possible, 
on directed networks backtracking may occur only
when the current node and the previous node are
connected in both directions.
Moreover, on a directed network an RW may become trapped in 
a dead-end node which does not have any outgoing links.
As a result, an RW path on a directed ER network may terminate
either by trapping or by retracing its path.
We obtain analytical results for the overall probabilities,
$p_{\rm trap}$ and $p_{\rm ret}$,
that an RW starting from a random node
will terminate by trapping or by retracing,
respectively.
It is found that in dilute networks most paths terminate
by trapping while in dense networks most paths 
terminate by retracing.
We obtain expressions 
for the
conditional probabilities, 
$P(d>\ell | {\rm trap})$ 
and
$P(d>\ell | {\rm ret})$,
of RWs which terminate by
trapping or by retracing, 
respectively,
as well for the conditional probabilities
$P({\rm trap} | d>\ell)$
and 
$P({\rm ret} | d>\ell)$.
It is found that the probability of termination by
trapping decreases with the path length while
the probability of termination by retracing increases
with the path length.
Since most of the directed networks in nature exhibit different distributions
of in-degrees and out-degrees, it is of interest to disentangle the effect
of these distributions on the distribution of first hitting times.
To this end, we extend our studies to a broader class
of directed networks, referred to as semi-ER networks,
in which the in-degree distribution is Poisson and the
out-degrees follow any desired distribution with the same mean as the 
in-degree distribution. 
We present analytical results for the distribution of first hitting times in such networks.
We also consider the distribution of last hitting times on directed ER networks.

The paper is organized as follows.
In Sec. 2 we present relevant properties 
of directed ER networks.
In Sec. 3 we describe the random walk model 
on directed ER networks.
In Sec. 4 we consider the 
evolution of the subnetwork of
the yet-unvisited nodes.
In Sec. 5 we present analytical results the distribution of last hitting times
of RWs on directed ER networks.
In Sec. 6 we present analytical results 
for the distribution of first hitting times
of RWs on directed ER networks.
In Sec. 7 we obtain analytical expressions for two central measures
(mean and median) and for a dispersion measure 
(the standard deviation) 
of this distribution.
In Sec. 8 we calculate the 
distributions of path lengths conditioned on the termination mechanism.
In Sec. 9 we generalize the analysis to directed semi-ER networks, which exhibit
a Poisson in-degree distribution and any desired out-degree distribution.
The results are summarized and discussed in Sec. 10.
In Appendix A we present the detailed calculation of the contribution
of the retracing mechanism to the distribution of first hitting times.

\section{The directed Erd\H{o}s-R\'enyi network}

The directed ER network is the simplest model of a directed random network.
It consists of $N$ nodes such that a directed edge, or link, 
from any node, $i$, to any other node, $j$, exists
with probability $p$, independently of 
the existence of the opposite link, or any other link.
Therefore, the probability that a random pair of nodes,
$i$ and $j$ are connected in both directions is $p^2$.
In directed networks each node, $i$ has an in-degree,
$k_i^{in}$, which is the number of incoming links and an out-degree,
$k_i^{out}$, which is the number of outgoing links. 
In general one should distinguish between the
degree distribution of incoming links, 
$p^{\rm in}(k)$, 
and the degree
distribution of outgoing links,
$p^{\rm out}(k)$
\cite{Havlin2010}. 
Since each link is directed out of one node
and into another node, the mean, 
$\langle k \rangle^{\rm in}$,
of
$p^{\rm in}(k)$
must be
equal to the mean,
$\langle k \rangle^{\rm out}$,
of
$p^{\rm out}(k)$.

In directed ER networks, 
both distributions,
$p^{\rm in}(k)$ 
and
$p^{\rm out}(k)$,
are 
binomial distributions. 
Therefore,
in the sparse limit
($p \ll 1$)
they are approximated by a 
Poisson distribution of the form 
\cite{Bollobas2001}

\begin{equation}
p^{\rm in}(k) = p^{\rm out}(k) = p(k)=\frac{{c}^{k}}{k!}e^{-c},
\label{eq:poisson}
\end{equation}

\noindent
where the mean in-degree and the mean out-degree are given by
$c=(N-1)p$.
It is important to note that the 
in-degree and out-degree of each node 
are uncorrelated.
Also, there are no degree-degree correlations between adjacent nodes.
The adjacency matrix, $A$, of a directed ER network
of $N$ nodes is a random $N \times N$ matrix 
with a zero diagonal and
whose off-diagonal
entries are $1$ with probability $p$ and $0$ with probability $1-p$.
Note that unlike the undirected case, $A$ is not necessarily a symmetric matrix.

The percolation properties of directed ER networks differ from those 
of their undirected counterparts
\cite{Havlin2010,Dorogovtsev2001,Newman2001}. 
Undirected ER networks exhibit a percolation
transition at $c=1$, above which the network consists of a giant cluster,
small, isolated components and isolated nodes. 
On the giant cluster, every
node can be reached from any other node. 
The percolation transition of a directed ER network also takes place at $c=1$,
above which a giant cluster is formed.
However, due to the directionality of the links, not all pairs of nodes on the giant
cluster can be reached from each other along paths which consist of directed links.
The subgraph of the giant cluster in which each pair of nodes can be reached from each other
in both directions is called the giant strongly connected component (GSCC)
\cite{Havlin2010,Dorogovtsev2001,Newman2001}.
The set of nodes on the giant cluster which can be reached from the 
GSCC
is called the out component,
while the set of nodes from which the 
GSCC
can be reached is called the in component. 
In addition, there are some nodes
on the giant cluster which do not belong either to the in component or to the
out component. These nodes are referred to as tendrils
\cite{Havlin2010}.
The probability of a random node in a directed ER network
to be an isolated node is $\exp(-2c)$. Also, the probability of a random
node to have only incoming links or only outgoing links is $\exp(-c)$.
A node which has only outgoing links cannot be reached unless it is
the initial node in the RW path. When the RW enters a node which has
only incoming links, it becomes trapped and the RW path terminates.

\section{The random walk model}

Consider an RW on a directed random network of $N$ nodes.
Each time step the walker chooses randomly one of the outgoing edges
of the current node, and hops along this edge to an adjacent node.
The RW path 
terminates when it steps into 
a node which it has already visited before
(retracing scenario) or when it enters a dead-end node which
has only incoming links (trapping scenario).
The initial node is chosen randomly among the
nodes for which the out-degree satisfies
$k_i^{\rm out} \ge 1$, so the RW is guaranteed to make
at least one move.
The resulting path length, $d$,
namely the number of steps 
pursued by the RW
until its termination, 
is referred to as the first hitting time.
In the analysis below we do not include the termination step itself
as a part of the RW path. 
This means that the path length of an RW which pursued
$\ell$ steps and terminated at the $\ell+1$ step
is $d=\ell$. 
The path includes $d+1$ nodes, 
since the initial node is also counted as
a part of the path.

\section{Evolution of the subnetwork of the yet-unvisited nodes}

Consider an RW starting from a random node 
on a directed ER network.
The RW divides the network into two subnetworks:
one consists of the already visited nodes and the
other consists of the yet-unvisited nodes. 
After $t$ time steps the size of the subnetwork of 
visited nodes is $t+1$ 
(including the initial node),
while the size of the network of
yet unvisited nodes is
$N(t)=N-t-1$.
Here we focus on the subnetwork of the yet-unvisited nodes. 
Its in-degree distribution and out-degree distribution evolve in time.
We denote these distributions,  at time $t$, by
$p^{\rm in}_t(k)$, 
and
$p^{\rm out}_t(k)$, $k=0,\dots,N(t)-1$,
respectively,
where
$p^{\rm in}_0(k)=p^{\rm out}_0(k) = p(k)$,
which is
given by Eq.
(\ref{eq:poisson}).
The mean in and out degrees of this subnetwork, 
which are given by

\begin{equation}
\langle k \rangle^{\rm in}_t 
= \sum_{k=0}^{N(t)-1} k p^{\rm in}_t(k),
\label{eq:<k>in}
\end{equation}

\noindent
and

\begin{equation}
\langle k \rangle^{\rm out}_t 
= \sum_{k=0}^{N(t)-1} k p^{\rm out}_t(k),
\label{eq:<k>out}
\end{equation}

\noindent
evolve accordingly.
Since the number of incoming links is equal to the number of outgoing links, 
the mean degrees of the two distributions must satisfy
$\langle k \rangle^{\rm in}_t = \langle k \rangle^{\rm out}_t$,
and are denoted by $c(t)$.

We now examine the evolution of the subnetwork of
the yet-unvisited nodes in terms of the mean 
numbers of incoming and outgoing links
which are removed at each step. 
Deleting a node along the RW path removes, on average, $c(t)$ incoming links
and $c(t)$ outgoing links from the node itself as well as $c(t)$ incoming
links and $c(t)$ outgoing links from its neighbors, which remain on
the subnetwork of the yet-unvisited nodes.
Denoting the initial in-degree and out-degree of node $i$, 
by 
$k^{\rm in}_i$
and
$k^{\rm out}_i$,
respectively
we note that 
$\sum_{i=1}^N k^{\rm in}_i = \sum_{i=1}^N k^{\rm out}_i = Nc$.
Thus, the time dependence of the mean degree can be expressed by

\begin{equation}
c(t) = \frac{Nc - 2\sum_{t^{\prime}=0}^{t-1} c(t^{\prime})  }{N-t}.
\end{equation}

\noindent
This implies that $c(t)$ obeys the recursion equation

\begin{equation}
c(t) = \left(1 - \frac{1}{N-t} \right) c(t-1),
\end{equation}

\noindent
in which the coefficient on the right hand side depends on $t$.
This equation is solved by

\begin{equation}
c(t) = \left( 1 - \frac{t}{N-1} \right) c.
\label{eq:coft}
\end{equation}

\noindent
For RWs on directed random networks,
there is a higher probability for the walker to
enter nodes with high incoming degrees. 
More precisely, 
the probability that in the next time step the RW will step into a node 
of in-degree $k$
is 
$k p^{\rm in}(k)/c$.
However, by the time the RW enters the next node, the previous
node is effectively deleted, together with the edge connecting the two nodes. 
Therefore, when the walker
enters a node of in-degree $k$, the in-degree 
of this node is reduced to $k - 1$.
%
%
A special property of the 
Poisson distribution is that
$k p(k) / c = p(k-1)$.
This means that
the probability that the node
visited at time $t+1$ will be of  
degree $k$ 
is given by 
$p_t(k-1)$. 
%
The outcome of this
reasoning is that the probability of the RW to visit a node 
of in-degree $k$ in the smaller network at time $t$ is simply
$p^{\rm in}_t(k)$, as if it makes a random 
choice of a node in the smaller network.
The result of this exact, yet delicate, balance is that
the subnetwork of the yet-unvisited nodes at time $t$,
is a directed ER network
with mean in and out degrees equal to $c(t)$,
as well as
degree distributions,
$p_t^{\rm in}(k)$
and
$p_t^{\rm out}(k)$,
given by

\begin{equation}
p_t(k) = \frac{c(t)^k}{k!}e^{-c(t)}.
\label{eq:p_t(k)}
\end{equation}

\noindent
It is interesting to note that a similar result is also obtained for 
undirected ER networks 
\cite{Tishby2016a}.

\section{The distribution of last hitting times}

The paths pursued by the RWs studied here are, in fact, segments of SAW paths.
In case of termination by trapping the RW path is identical to an SAW path, while
in case of termination by retracing the RW path consists of the
initial segment of a longer SAW path.
Therefore, the distribution of first hitting times is bounded from above
by the distribution of path lengths of SAWs on the same networks,
also referred to as the last hitting times.
Below we present analytical results for the distribution of last hitting times.
Consider an SAW on an ER network, 
which starts from a random node, $i$, with an out-degree
$k^{\rm out}_i \ge 1$.
The SAW hops through directed links 
between adjacent nodes until it reaches a node
from which it cannot exit.
At that stage the path terminates. 
The out-degree of node $i$ in the subnetwork of the
yet-unvisited nodes, at time $t$,
is given by 
$k_i^{\rm out}(t)$.
The termination of an SAW path occurs when it enters
a node for which
$k_i^{\rm out}(t) = 0$.
The probability that a random node 
does not have outgoing links in the subnetwork
of the yet-unvisited nodes at time $t$ is
$p_t^{\rm out}(k=0)$. 
The conditional probability that the SAW will proceed from time
$t$ to time $t+1$ without being trapped is denoted
by $P(d>t|d>t-1)$, where $d$ represents the random variable
of the path length and $t$ represents its actual value.
This conditional probability is given by
$P(d > t|d > t-1) = 1-p_t^{\rm out}(k=0)$.
Thus, the probability that the path length of the SAW will
be longer than $\ell$, 
also known as the tail distribution,
is given by

\begin{equation} 
P(d>\ell) = P(d>0) \prod_{t=1}^{\ell} P(d > t|d > t-1).
\label{eq:cond}
\end{equation}

\noindent
The probability $P(d>0)=1$
since the initial node is chosen randomly among the nodes for which
$k_i^{\rm out} \ge 1$.
Thus, the tail distribution takes the form

\begin{equation} 
P(d>\ell) = \prod_{t=1}^{\ell} \left[1-p_{t}^{\rm out}(k=0) \right].
\label{eq:cond2L}
\end{equation}

\noindent
While Eq.
(\ref{eq:cond2L})
applies to any network, 
in the case of a directed ER network there is an explicit expression for
the probability of a node to have no outgoing links at time $t$, of the form
$p_t^{\rm out}(k=0)=\exp[-c(t)]$.
Therefore, the tail distribution takes the form

\begin{equation}
P(d>\ell)
=\prod_{t=1}^{\ell} \left[1-e^{-c(t)}\right].
\label{eq:cond3}
\end{equation}

\noindent
To obtain a closed form expression for the tail distribution, 
$P(d>\ell)$,
we
take the natural 
logarithm on both sides of Eq.
(\ref{eq:cond3}).
This leads to

\begin{equation}
\ln \left[P\left(d>\ell\right)\right] =
\sum_{t=1}^{\ell} \ln \left[1-\exp\left(\frac{ct}{N-1}-c\right) \right].
\label{eq:lht_sum}
\end{equation}

\noindent
Replacing this sum by an integral we obtain

\begin{equation}
\ln \left[P\left(d>\ell\right)\right]
=
\int_{1/2}^{\ell+1/2}\ln 
\left[1-\exp \left( \frac{ct}{N-1}-c \right) \right]dt,
\label{eq:lht_int}
\end{equation}

\noindent
where the limits of the integration are set
such that the summation over each integer, $i$, is replaced by an
integral over the range $(i-1/2,i+1/2)$.
This integral is in fact a partial Bose-Einstein integral, which 
can be expressed in terms of
the Polylogarithm ${\rm Li}_{n}(x)$
function
\cite{Olver2010}

\begin{equation}
P(d>\ell)
=
\exp \left\{\frac{N-1}{c}\left[{\rm Li}_{2}
\left(e^{- \left(1- \frac{1}{2(N-1)}\right) c}
\right)-{\rm Li}_{2}\left(e^{- \left(1- \frac{\ell+1/2}{N-1} \right) c}
\right)\right]\right\}.
\label{eq:P(d>l)}
\end{equation}

\noindent
In approximating the sum of 
Eq. (\ref{eq:lht_sum}) by the integral of Eq. (\ref{eq:lht_int})
we have used the formulation of the middle Riemann sum. 
Since the function
$\ln[P(d>\ell)]$ is a monotonically decreasing function, the value of
the integral is over-estimated by the left Riemann sum, $L(\ell)$, and under-estimated
by the right Riemann sum, $R(\ell)$. 
The error involved in this approximation is thus 
bounded by the difference $\Delta(\ell) = L(\ell) - R(\ell)$,
which satisfies
$\Delta(\ell) = \ln(e^c - 1) - \ln [e^c - e^{c (\ell+1)/(N-1)}]$.
Thus, the relative error in $P(d>\ell)$ due to the approximation of the sum
by an integral is bounded by
$\eta_{\rm SI} \sim \ell/N$.
Comparing the values obtained from the sum and the integral
over a broad range of parameters, we find that the pre-factor of
the error is very small, so in practice the error introduced by approximation of the
sum by an integral is negligible.

For large networks ($N \gg 1$)
one can further approximate the ${\rm Li}_2(x)$ function in
Eq.
(\ref{eq:P(d>l)})
by the leading term in its Taylor expansion of the form
${\rm Li}_2(x) = \sum_{k=1}^{\infty} x^k/k^2$.
We obtain

\begin{equation}
P\left(d>\ell\right)
\simeq
\exp\left[-\frac{N}{c}e^{-c}
\left(e^{\frac{c}{N}\ell}-1\right)\right].
\label{eq:tail2}
\end{equation}

\noindent
Evaluating the second order term we find that the relative
error involved in this approximation is
$\eta_{TE} \sim \ell/e^{2c}$.
The expression for the tail
distribution of last hitting times for directed ER networks, presented in Eq.
(\ref{eq:tail2})
coincides with the  
Gompertz distribution 
\cite{Gompertz1825,Johnson1995,Shklovskii2005,Ohishi2008}
of a random variable $X$,
which takes the form

\begin{equation}
P(X>x) =
\exp\left[-\eta
\left(e^{ax}-1\right)\right] 
\label{eq:Gompertz}
\end{equation}

\noindent
for $x \ge 0$,
with the scale parameter
$a = {c}/{N}$
and the shape parameter 
$\eta={N}e^{-c}/c$.

\section{The distribution of first hitting times}

An RW on a directed network hops randomly along directed edges
until it steps into a previously visited node (retracing) or gets trapped in a node
which does not have any outgoing edges (trapping).
Unlike the case of an undirected network, a directed edge from the current
node to the previous node exists only with probability $p$.
Therefore, on a directed network the probability of a 'backtracking' move 
of the RW from the current node to the previous node is the same as
the probability to hop into any other node. 
As a result, on a directed network
one does not need to treat the backtracking step as a special termination
scenario which is different from the retracing scenario.
The RW path on a directed network may terminate by retracing starting from
the second time step, by hopping back into the initial node (in case that
such a backward link exists, which occurs with probability $p$).
However, as
the termination move is not counted, the resulting path length
is $\ell=1$.
Since the initial node is chosen among the nodes which have at least
one outgoing edge, termination by trapping cannot occur in the first step,
justifying the statement that
it may occur starting only from the second time step.

We denote the conditional probability that the RW path 
exceeds $t$ steps, given that it exceeds $t-1$ steps, by
$P(d>t | d>t-1)$.
It can be expressed in the form

\begin{equation}
P(d>t | d>t-1) 
= P_{\rm ret}(d>t | d>t-1) 
P_{\rm trap}(d>t | d>t-1),
\end{equation}

\noindent
where
$P_{\rm trap}(d>t | d>t-1)$ 
is the probability that the RW path will not terminate by trapping at the
$t+1$ time step, while
$P_{\rm ret}(d>t | d>t-1)$
is the probability that it will not terminate by retracing, given 
that it has not terminated by trapping.

The probability that an RW will not terminate by trapping 
in the $t+1$ time step is given by the probability that the
node it entered 
at time $t$
has an out-degree $k_{out} > 0$
in the entire network
(namely counting all its outgoing links regardless of whether they point
towards nodes which were already visited or yet unvisited).
This probability is given by

\begin{equation}
P_{\rm trap}(d > t|d > t-1) =
\sum_{k=1}^{N-1} p^{\rm out}(k) = 1 - p^{\rm out}(0).
\label{eq:trap2}
\end{equation}

\noindent
Inserting in 
Eq. 
(\ref{eq:trap2})
the Poisson distribution 
of Eq.
(\ref{eq:poisson})
we obtain 

\begin{equation}
P_{\rm trap}(d > t|d > t-1) = 1 - e^{-c}.
\label{eq:trap3}
\end{equation}

\noindent
Note that this probability does not depend on the time, $t$.

Given that the RW has not terminated by trapping at the
$t+1$ time step, the out-degree of the current node 
at time $t$ is conditioned to be $k^{\rm out} > 0$.
We will now evaluate the probability,
$P_{\rm ret}(d > t|d > t-1)$, 
that the RW will also not terminate by retracing.
This probability is given by
$P_{\rm ret}(d > t|d > t-1) = \langle k^{\rm out}(t) | k^{\rm out}>0 \rangle /
\langle k^{\rm out} | k^{\rm out} > 0 \rangle$,
where $k^{\rm out}(t)$ is the out-degree of the current node within the sub-network 
of the yet-unvisited nodes.
Putting aside for a moment the condition
$k^{\rm out} > 0$,
there are $N-1$
nodes in the network which may
receive an outgoing link from 
the current node, 
each one of them with probability $p$.
Thus, the expectation value of the number of outgoing links
from the current node to its neighbors, 
through which the RW may hop at the $t+1$ time step, 
is $c=(N-1)p$.
Since the number of yet unvisited nodes is $N-t-1$, we conclude that
the current node is expected to have
$c(t) = (N-t-1)p$ neighbors which have not yet been visited.
In presence of the condition
$k^{\rm out} > 0$,
we find that
$\langle k^{\rm out}(t) | k^{\rm out}>0 \rangle = c(t)/(1-e^{-c})$
and
$\langle k^{\rm out} | k^{\rm out} > 0 \rangle = c/(1-e^{-c})$.
We thus obtain
%
%

\begin{equation} 
P_{\rm ret}(d>t|d>t-1) = \frac{c(t)}{c}. 
\label{eq:P_r2}
\end{equation}

\noindent
Combining the results presented above, it is found that
the probability that the path of the RW will not terminate 
at the $t+1$ time step
is given by the conditional probability

\begin{equation} 
P(d > t|d > t-1) = 
\frac{c(t)}{c} 
\left(1 - e^{-c} \right).
\label{eq:cond3p}
\end{equation}

\begin{figure}
\centerline{
\includegraphics[width=8cm]{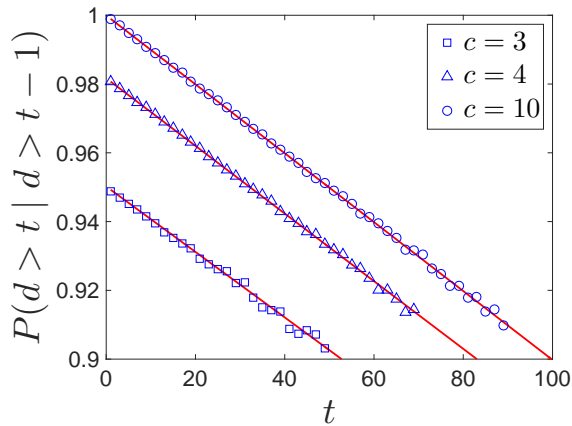}
}
\caption{
The conditional probability
of first hitting times, 
$P(d > t|d > t-1)$ 
vs. $t$,
obtained from Eq.
(\ref{eq:cond3p})
(solid lines)
and 
from numerical simulations of RWs 
(symbols)
on directed ER networks of
size $N=1000$ and mean degrees $c=3$, $4$ and $10$ 
(squares, triangles and circles, respectively).
The analytical and numerical results are found to
be in very good agreement.
} 
\label{fig:1}
\end{figure}

\noindent
In Fig. 
\ref{fig:1} 
we present
the conditional probability 
$P(d > t|d > t-1)$ 
of first hitting times
vs. $t$ for an RW on a directed ER 
network 
of size $N=1000$ and three values of $c$. 
The analytical results 
(solid lines) 
obtained from Eq.
(\ref{eq:cond3p})
are found to be in very good agreement with the numerical simulations
(symbols),
confirming the validity of this equation.
Note that the numerical results become more noisy as $t$
increases, due to diminishing statistics, 
and eventually terminate.
This is particularly apparent for the 
smaller values of $c$.

The 
tail distribution, $P(d>\ell)$,
namely the
probability that the path length of the RW will
be longer than 
$\ell$ is given by

\begin{equation} 
P(d>\ell) = P(d>0) \prod_{t=1}^{\ell} P(d > t|d > t-1).
\label{eq:cond1}
\end{equation}

\noindent
The probability $P(d>0)=1$,
since the initial node is chosen among
the nodes with out-degrees $k_{\rm out} \ge 1$.
The probability 
$P(d>\ell)$ 
can be written as a product of the form

\begin{equation} 
P(d>\ell) =  P_{\rm ret}(d>\ell) P_{\rm trap}(d>\ell),
\label{eq:cond2}
\end{equation}

\noindent
where

\begin{equation} 
P_{\rm trap}(d>\ell) =  \prod_{t=1}^{\ell} 
\left( 1 - e^{-c} \right),
\label{eq:cond4}
\end{equation}

\noindent
and

\begin{equation} 
P_{\rm ret}(d>\ell) =  \prod_{t=1}^{\ell} 
\frac{c(t)}{c}.
\label{eq:cond6}
\end{equation}

\noindent
To obtain a closed form expression for the tail distribution, 
we
take the natural 
logarithm on both sides of Eq.
(\ref{eq:cond2}).
This leads to

\begin{equation}
\ln \left[P\left(d>\ell\right)\right] =
\ln \left[P_{\rm ret}\left(d>\ell\right)\right]+
\ln \left[P_{\rm trap}\left(d>\ell\right)\right].
\label{eq:logtail1}
\end{equation}

\noindent
The calculation of 
$P_{\rm trap}(d>\ell)$ 
is simplified by the fact that 
$P_{\rm trap}(d>t | d>t-1)$
does not depend on $t$.
As a result, Eq.
(\ref{eq:cond4})
can be written in the form

\begin{equation}
P_{\rm trap}(d>\ell) 
= \left(1 - e^{-c} \right)^{\ell}.
\label{eq:tailb}
\end{equation}

%
%
%

\noindent
The termination by the trapping scenario can thus be considered as a 
Poisson process, in which
the termination probability  
is fixed and depends only on the mean
degree of the network.
Taking the logarithm on both sides of Eq.
(\ref{eq:tailb})
we obtain

\begin{equation}
\ln \left[P_{\rm trap}\left(d>\ell\right)\right] 
= \ell \ln \left(1 - e^{-c} \right).
\end{equation}

\noindent
The detailed evaluation of $P_{\rm ret}(d>\ell)$ is presented in Appendix A,
where $\ln P_{\rm ret}(d>\ell)$ is expressed as a sum, the sum is replaced
by an integral and the integration is performed.
Combining the results for
$P_{\rm ret}\left(d>\ell\right)$
and 
$P_{\rm trap}\left(d>\ell\right)$
we obtain the tail distribution

\begin{equation}
\fl
P\left(d>\ell\right)
\simeq
\exp\left[\left(\ell+\frac{3}{2}-N\right)
\ln\left(1-\frac{\ell+1/2}{N-1}\right)-\ell
+\ell \ln \left( 1 - e^{-c} \right) -\frac{1}{2}\right].
\label{eq:tail_noapp}
\end{equation}

\noindent
Assuming that the RW paths are short compared to the network size, 
namely that
$\ell \ll N$,
one can use the approximation

\begin{equation}
\ln\left(1-\frac{\ell+1/2}{N-1}\right)
\simeq
-\frac{\ell+1/2}{N-1} - 
\frac{(\ell+1/2)^2}{2 (N-1)^2},
\label{eq:lnapprox}
\end{equation}

\noindent
which yields

\begin{equation}
P\left(d>\ell\right)
\simeq
\exp\left[-\frac{(\ell+1/2)^{2}}{2N}
+\ell\ln\left(1-{e^{-c}}\right)\right].
\end{equation}

\noindent
Thus, the tail distribution of first hitting times of RWs
on directed ER networks
takes the form

\begin{equation}
P\left(d>\ell\right)
\simeq
\exp\left[-\left(\frac{\ell}{2\alpha}\right)^{2}-\beta\ell\right]
\label{eq:tail_simp}
\end{equation}

\noindent
where

\begin{equation}
\alpha=\sqrt{\frac{N}{2}}
\label{eq:alphaN2}
\end{equation}

\noindent
and

\begin{equation}
\beta = -\ln(1 - e^{-c}) 
+ \frac{1}{4 \alpha^2}.
\end{equation}

\noindent
The right hand side of Eq. 
(\ref{eq:tail_simp})
is, in fact, a product of a 
discrete Rayleigh distribution 
\cite{Olver2010}
and a discrete exponential distribution.
The Rayleigh distribution accounts for the retracing
scenario while the exponential distribution accounts for the trapping scenario.

Considering the next order in the series expansion of Eq.
(\ref{eq:lnapprox}),
we find that the relative error in 
Eq. (\ref{eq:tail_simp}),
for $P(d>\ell)$,
due to the truncation of the Taylor expansion after the second order
is $\eta_{\rm TE} \sim  \ell^3/N^2$.
This error is very small as long as $\ell \ll N^{1/2}$.
Note that paths of length $\ell \simeq N^{1/2}$,
for which the error 
in $P(d>\ell)$
is noticeable, become prevalent only
in the limit of dense networks, where $c > N^{1/2}$.

\begin{figure}
\centerline{
\includegraphics[width=15cm]{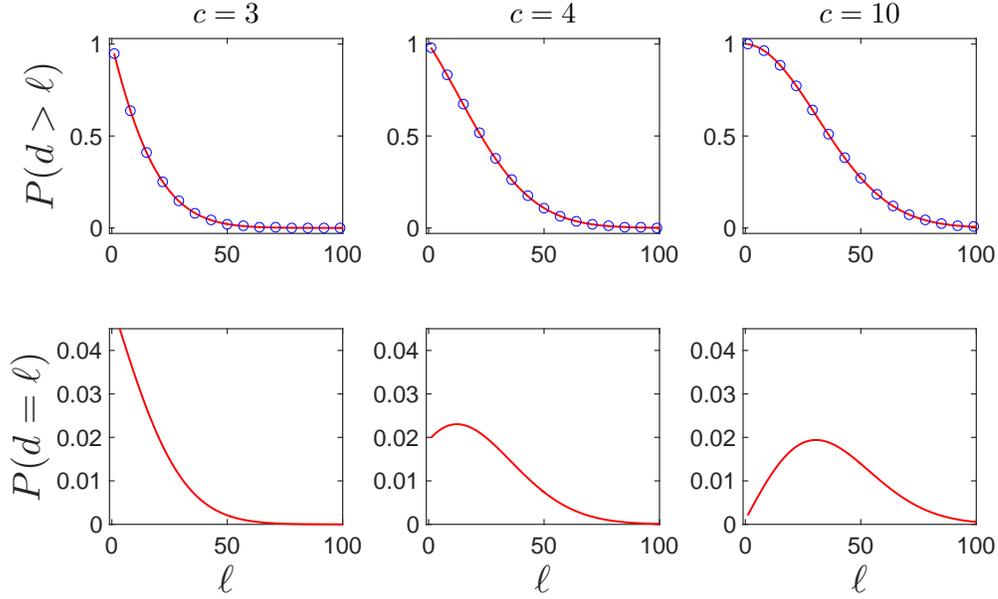}
}
\caption{
The tail distributions 
(top row)
of first hitting times, 
$P(d > \ell)$ vs. $\ell$,
of RWs on directed ER networks
of size $N=1000$ 
and 
$c=3$, $4$ and $10$.
The theoretical results, obtained from Eq.
(\ref{eq:tail_noapp}) 
(solid lines)
are in excellent agreement with 
the results obtained
from numerical simulations (circles).
The corresponding probability distribution functions,
$P(d=\ell)$,
obtained from Eq.  
(\ref{eq:pdf_simp}),
are shown in the bottom row. The agreement with the
numerical results is already established in the top row and therefore the numerical
data is not shown in the bottom row.
}
\label{fig:2}
\end{figure}

\noindent
In Fig.
\ref{fig:2}
we present the
tail distributions 
(top row)
of first hitting times, 
$P(d > \ell)$,
of RWs on directed ER networks
of size $N=1000$ 
and 
$c=3$, $4$ and $10$.
The theoretical results
(solid lines),
obtained from Eq.
(\ref{eq:tail_noapp}), 
are in excellent agreement with 
the results obtained
from numerical simulations (circles).
This agreement indicates that the approximation used in the
analytical derivation, namely the replacement of a sum by an
integral, is justified.
The corresponding probability density functions,

\begin{equation}
P(d=\ell) = P(d>\ell-1) - P(d>\ell),
\label{eq:pdf_simp}
\end{equation}

\noindent
are shown in the bottom row.
It is found that for small values of $c$ most paths are short
and the probability density function is a monotonically
decreasing function of $\ell$.
As $c$ is increased, the distribution
$P(d=\ell)$
shifts to the right and develops a well defined peak.

\section{Central and dispersion measures}

In order to characterize the distribution of 
first hitting times of RWs on directed ER networks we
derive expressions for the mean, median and standard deviation of 
this distribution.
The mean of the distribution can be obtained
from the tail-sum formula
\cite{Pitman1993}

\begin{equation}
\ell_{\rm mean}(N,c) =
\sum_{\ell=0}^{N-2} P(d>\ell). 
\label{eq:tailsum1}
\end{equation}

\noindent
Assuming that the 
out-degree of the
initial node satisfies $k_{\rm out} \ge 1$,
this sum can be written in the form

\begin{equation}
\ell_{\rm mean}(N,c) =
1 + \sum_{\ell=1}^{N-2} P(d>\ell). 
\label{eq:tailsum2}
\end{equation}

\noindent
Expressing the sum 
as an integral we obtain

\begin{equation}
\ell_{\rm mean}(N,c) \simeq
1 + \int_{\frac{1}{2}}^{N-\frac{3}{2}} P(d>\ell)d\ell.
\label{eq:ell_mean1}
\end{equation}

\noindent
Inserting 
$P(d>\ell)$
from Eq.
(\ref{eq:tail_simp})
we obtain

\begin{equation}
\ell_{\rm mean}
\simeq
1+\int_{\frac{1}{2}}^{N-\frac{3}{2}}
\exp\left[-\left(\frac{\ell}{2\alpha}\right)^{2}
-\beta\ell\right] d\ell.
\label{eq:ell_mean3}
\end{equation}

\begin{figure}
\centerline{
\includegraphics[width=19cm]{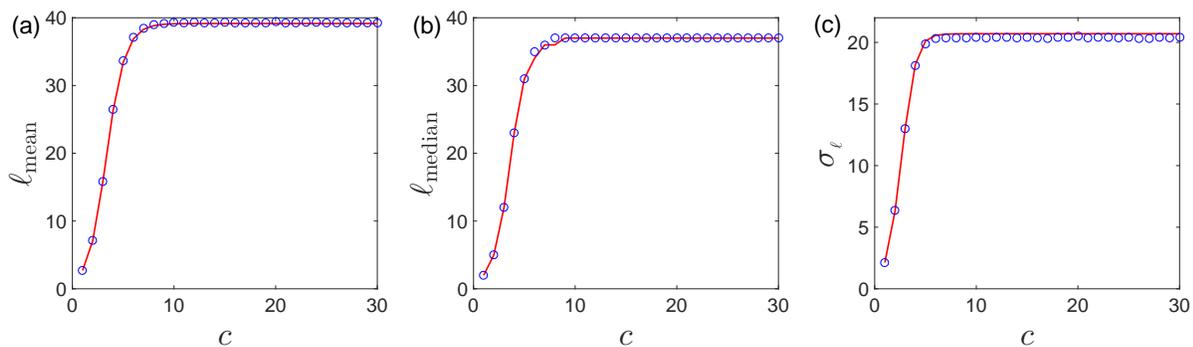}
}
\caption{
The mean, $\ell_{\rm mean}$ (a), median, $\ell_{\rm median}$ (b) 
and the standard deviation $\sigma_{\ell}$ (c),
of the distribution of first hitting times
of RWs on directed ER networks of size $N=1000$,
as a function of the mean degree,
$c$.
The analytical results (solid lines), 
obtained from Eqs.
(\ref{eq:ell_mean}),
(\ref{eq:ell_median})
and
(\ref{eq:sigma_ell}),
respectively,
are in excellent agreement with numerical 
simulations 
(circles). 
}
\label{fig:3}
\end{figure}

\noindent
Solving the integral for $N \gg 1$ we obtain

\begin{equation}
\ell_{mean}
\simeq
1+\sqrt{\pi}\alpha e^{\alpha^{2}\beta^{2}}
\left[1-{\rm erf}\left(\alpha\beta+\frac{1}{4\alpha}\right)\right],
\label{eq:ell_mean}
\end{equation}

\noindent
where ${\rm erf}(x)$ is the error function, also called the Gauss error function
\cite{Olver2010}.
This function exhibits a sigmoid shape. For 
$|x| \ll 1$ 
it can be
approximated by 
${\rm erf}(x) \simeq 2x/\sqrt{\pi}$ 
while for 
$|x| > 1$ 
it quickly converges to 
${\rm erf}(x) \rightarrow {\rm sign}(x)$.
For small values of $c$, the mean path lengths $\ell_{\rm mean}$
quickly increases as $c$ is increased, until it saturates.
The saturation is obtained at 
$c \simeq (\ln N)/2$.
The saturation value of $\ell_{\rm mean}$
is
$\ell_{\rm mean} \simeq 1 + \sqrt{ {\pi N}/{2} }$.
In Fig.
\ref{fig:3}(a)
we present analytical results (solid lines) for the
mean,
$\ell_{\rm mean}$,
of the distribution of first hitting times
of RWs on directed ER networks of size $N=1000$,
as a function of the mean degree $c$.
The analytical results
are found to be in very good agreement with the results of
numerical simulations (circles).

To obtain a more complete characterization of the distribution of first
hitting times, it is also useful to evaluate its median, $\ell_{\rm median}$.
Here the
median is defined as the value of $\ell$ for which
$| P(d>\ell) - P(d<\ell) | \rightarrow {\rm min}$,
where $\ell$ may take either an integer or a half-integer value.
For integer values of $\ell$, $P(d<\ell) = 1 - P(d>\ell-1)$
while for half integar values of $\ell$, $P(d<\ell) = 1- P(d>\ell)$.
Expressing $P(d>\ell)$ by Eq.
(\ref{eq:tail_simp}),
we find that 
$\ell_{\rm median}$
can be approximated by

\begin{equation}
\ell_{median}
\simeq
\frac{1}{2} 
\left\lfloor 4 \alpha \sqrt{ \alpha^2 \beta^2 + \ln 2 } - 4 \alpha^2 \beta + 1/2 \right\rfloor,
\label{eq:ell_median}
\end{equation}

\noindent
where $\lfloor x \rfloor$ is the largest integer which is smaller than $x$.
For small values of $c$, the median, $\ell_{\rm median}$
quickly increases as $c$ is increased, until it saturates.
The saturation is obtained at 
$c \simeq (\ln N)/2$.
The saturation value of $\ell_{\rm median}$
is
$\ell_{\rm median} \simeq  \lceil \sqrt{ 2 N \ln 2} \rceil$,
which is slightly lower than the saturation value of $\ell_{\rm mean}$.
In Fig.
\ref{fig:3}(b)
we present analytical results (solid lines) for the
median, 
$\ell_{\rm median}$, 
as a function of the mean degree $c$.
The analytical results
are found to be in very good agreement with the results of
numerical simulations (circles).

The moments of the distribution of RW path lengths,
$\langle \ell^n \rangle$,
are given by
the tail-sum formula
\cite{Pitman1993}

\begin{equation}
\langle \ell^n \rangle = 
\sum_{\ell=0}^{N-2} [(\ell+1)^n - \ell^n] P(d>\ell).
\label{eq:tail_sumn}
\end{equation}

\noindent
Using this formula to evaluate the second moment 
and replacing the sum by an integral we obtain

\begin{equation}
\langle \ell^{2} \rangle = 1 +
\int_{\frac{1}{2}}^{N-\frac{3}{2}}\left(2\ell+1\right)
\exp\left[-\left( \frac{\ell}{2 \alpha} \right)^{2}
-\beta\ell\right] d\ell.
\label{eq:ell2}
\end{equation}

\noindent
Solving the integral and taking the limit $N \gg 1$,
we obtain

\begin{equation}
\fl
\langle \ell^{2} \rangle =
1
+\alpha e^{\alpha^2 \beta^2}
\left\{ \sqrt{\pi} \left(1-4\alpha^{2} \beta \right)
\left[1- {\rm erf}\left(\frac{1+4\alpha^2\beta}{4 \alpha}\right)\right]
+4\alpha e^{-\left(\frac{1+4\alpha^2\beta}{4\alpha}\right)^{2}}\right\}.
\end{equation}

\noindent
The standard deviation 
$\sigma_{\ell}(c)$
is given by

\begin{equation}
\sigma_{\ell}^2(c) = \langle \ell^{2} \rangle - \ell_{mean}^{2}.
\label{eq:sigma_ell}
\end{equation}

\noindent
For small values of $c$, the standard deviation, $\sigma_{\ell}$,
quickly increases as $c$ is increased, until it saturates.
The saturation level of the standard deviation is
$\sigma_{\ell} \simeq \sqrt{(4-\pi)N/2}$.
In Fig.
\ref{fig:3}(c)
we present analytical results (solid lines) for the
standard deviation,
$\sigma_{\ell}$,
as a function of the mean degree $c$.
The analytical results
are found to be in very good agreement with the results of
numerical simulations (circles).

\section{Analysis of the two termination mechanisms}

The path of an RW on a directed ER network
may terminate either by the
trapping scenario or by the retracing scenario.
Since the initial node is chosen such that its out-degree
satisfies
$k_{out} \ge 1$,
the trapping mechanism may occur starting from the
second step of the RW. 
The probability of trapping 
is 
$\exp(-c)$ 
at any time step afterwards. 
The termination by retracing takes place when the RW steps into
a node which it has already visited before.
This may also occur starting from
the second time step of the RW. 
The probability that an RW will terminate
by retracing increases with time.
This is due to the fact that each visited node
becomes a potential termination site.
In the limit of sparse networks, 
paths which terminate after a small number of
steps are likely terminate by trapping, while paths
which survive for a long time are more likely to terminate by
retracing. 
In denser networks the probability of termination by
trapping is much lower than the probability of termination
by retracing even for short paths.
Below we present a detailed analysis of the probabilities
of an RW to terminate by trapping or by retracing. 
We denote by $p_{\rm trap}$
the probability that an RW 
starting from a random initial node
will eventually terminate
by the trapping scenario
and by $p_{\rm ret}$ 
the probability that it will terminate
by the retracing scenario.
These two probabilities satisfy
$p_{\rm trap} + p_{\rm ret} = 1$.

It is of interest to study 
the conditional distributions,
$P(d=\ell | {\rm trap})$, of 
paths terminated by trapping, 
and
$P(d=\ell | {\rm ret})$, 
of paths terminated by retracing.
These distributions satisfy the normalization 
conditions
$\sum_{\ell=2}^{N-1} P(d= \ell | {\rm trap}) =1$
and
$\sum_{\ell=2}^{N-1} P(d= \ell | {\rm ret}) =1$.
The overall distribution of path lengths can be decomposed in terms of 
the conditional distributions according to

\begin{equation}  
P(d=\ell) = p_{\rm trap} P(d=\ell | {\rm trap}) + p_{\rm ret} P(d=\ell | {\rm ret}).
\label{eq:br}
\end{equation}

\noindent
The first term on the right hand side of Eq.
(\ref{eq:br}) 
can be written as

\begin{equation}
p_{\rm trap} P(d=\ell | {\rm trap}) =
P(d>\ell-1) \left[1-P_{\rm trap}(d>\ell|d>\ell-1) \right],
\label{eq:p_b}
\end{equation}

\noindent
namely as the probability that the RW will pursue $\ell$ steps
and will terminate at the $\ell+1$ step by the trapping scenario.
The second term on the right hand side of Eq.
(\ref{eq:br}) 
can be written as

\begin{equation}
\fl
p_{\rm ret} P(d=\ell | {\rm ret}) =
P(d>\ell-1) P_{\rm trap}(d>\ell|d>\ell-1)
\left[1-P_{\rm ret}(d>\ell|d>\ell-1)\right],
\label{eq:p_r}
\end{equation}

\noindent
namely as the probability that the RW will pursue $\ell$ steps,
then in the $\ell+1$ step it will not get trapped but will
retrace its path by stepping into a node which 
was already visited before.
Summing up both sides of Eq.
(\ref{eq:p_b})
over all integer values of $\ell$ we conclude that

\begin{equation}
p_{\rm trap} = e^{-c}  \sum_{\ell=1}^{N-1} P(d>\ell-1).
\end{equation}

\noindent
Using the tail-sum formula, 
Eq. 
(\ref{eq:tailsum1}),
we find that the
probability that the RW will terminate 
by the trapping scenario is actually

\begin{equation}
p_{\rm trap} = e^{-c} \ell_{\rm mean}.
\label{eq:p_b2}
\end{equation}

\noindent
Therefore, the probability of the RW to terminate by
retracing is

\begin{equation}
p_{\rm ret} = 1 - e^{-c} 
\ell_{\rm mean}.
\label{eq:p_r2}
\end{equation}

\begin{figure}
\centerline{
\includegraphics[width=8cm]{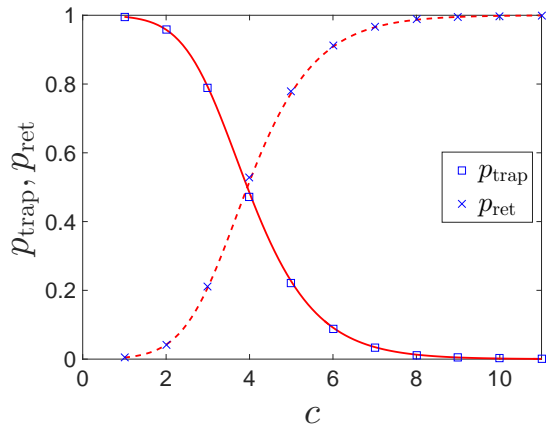}
}
\caption{
The probabilities 
$p_{\rm trap}$ 
and 
$p_{\rm ret}$
that an RW on a directed ER network will terminate by trapping or
by retracing, respectively,
as a function of the mean degree, $c$.
The theoretical results, obtained from Eqs.
(\ref{eq:p_b2})
and
(\ref{eq:p_r2})
are found to be in excellent agreement with 
the results of numerical simulations 
(symbols).
}
\label{fig:4}
\end{figure}

\noindent
In Fig.
\ref{fig:4}
we present the probability $p_{\rm trap}$ that an RW
on a directed ER network of size $N=1000$ will terminate by trapping
and the probability $p_{\rm ret}$ that it will terminate by retracing,
as a function
of the mean degree, $c$.
As expected, the probability $p_{\rm trap}$ is a decreasing function of $c$ 
while $p_{\rm ret}$
is an increasing function of $c$.

Using Eq. 
(\ref{eq:p_b})
the conditional probability
$P(d=\ell | {\rm trap})$ 
can be written in the form

\begin{equation}
P(d=\ell | {\rm trap}) = \frac{P(d>\ell-1)}{\ell_{\rm mean}},
\label{eq:p_lbeq}
\end{equation}

\noindent
where
$P(d>\ell-1)$
is given by Eq.
(\ref{eq:tail_simp}).
Similarly, the conditional probability
$P(d=\ell | {\rm ret})$
takes the form

\begin{equation}
P(d=\ell | {\rm ret}) =  \left( \frac{ 1-e^{-c} }{c} \right) 
\left[ \frac{ c -c(\ell) }{1 - e^{-c} \ell_{\rm mean}} \right] P(d>\ell-1), 
\label{eq:p_lreq}
\end{equation}

\noindent
where $c(\ell)$ is given by Eq.
(\ref{eq:coft}).
The corresponding tail distributions
take the form

\begin{equation}
P(d > \ell | {\rm trap}) = 
\frac{\sum_{t=\ell+1}^{N-1} P(d>t-1)}{\ell_{\rm mean}},
\label{eq:p_lbgt}
\end{equation}

\noindent
and

\begin{equation}
P(d > \ell | {\rm ret}) =  \left( \frac{1-e^{-c}}{c} \right) 
\sum_{t=\ell+1}^{N-1} 
\left[ \frac{c - c(t)}{1-e^{-c}\ell_{\rm mean}} \right] P(d>t-1).
\label{eq:p_lrgt}
\end{equation}

\noindent
Replacing the sums in Eqs.
(\ref{eq:p_lbgt})
and
(\ref{eq:p_lrgt})
by integrals and carrying out the integrations,
we obtain

\begin{equation}
P(d > \ell | {\rm trap}) 
\simeq
\frac{ \sqrt{\pi} \alpha e^{\alpha^2 \beta^2} 
\left[ 1 - {\rm erf} \left( \frac{2 \alpha^2 \beta + \ell - 1/2}{2 \alpha} \right) \right] }
{1 + \sqrt{\pi} \alpha e^{\alpha^2 \beta^2} 
\left[ 1 - {\rm erf} \left( \frac{4 \alpha^2 \beta + 1}{4 \alpha} \right) \right] },
\label{eq:p_lbgti}
\end{equation}

\noindent
and

\begin{eqnarray}
P(d > \ell | {\rm ret}) 
&\simeq&
\left[ \frac{1-e^{-c}}{\left(1 - e^{-c} \ell_{\rm mean} \right) N }\right]
\left\{ 2 \alpha^2 e^{ - \frac{(\ell-1/2)^2}{4 \alpha^2} -\beta (\ell-1/2) } \right.
\nonumber \\
&-& \left. \sqrt{\pi} \alpha (2 \alpha^2 \beta - 1) e^{\alpha^2 \beta^2}
\left[ 1 - {\rm erf}\left( \frac{2 \alpha^2 \beta + \ell - 1/2}{2 \alpha} \right) \right] \right\}.
\label{eq:p_lrgti}
\end{eqnarray}

\begin{figure}
\centerline{
\includegraphics[width=15cm]{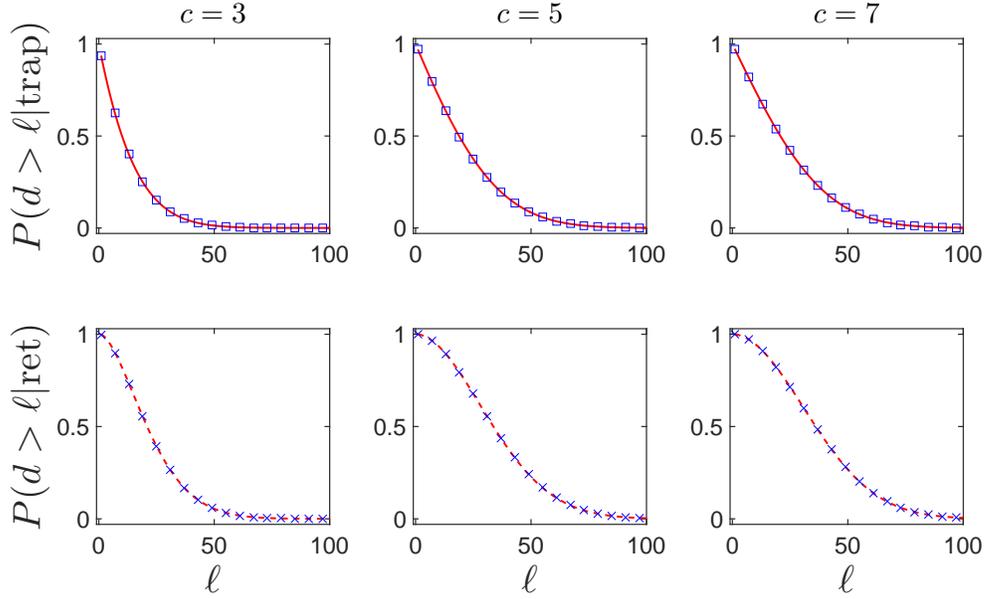}
}
\caption{
The conditional tail distributions 
$P(d>\ell | {\rm trap})$
and
$P(d>\ell | {\rm ret})$
of first hitting times 
vs. $\ell$
for RWs on a directed ER network, 
for paths terminated by trapping
(top row) or by retracing (bottom row),
respectively.
The results are shown for $N=1000$
and $c=3$, $5$ and $7$.
The theoretical results for 
$P(d>\ell | {\rm trap})$ 
are obtained from Eq.
(\ref{eq:p_lbgt}),
while the theoretical results for
$P(d>\ell | {\rm ret})$ 
are obtained from Eq.
(\ref{eq:p_lrgt}).
In both cases, the theoretical results (solid lines) are
found to be in excellent agreement with the numerical simulations
(symbols).
}
\label{fig:5}
\end{figure}

\noindent
In Fig.
\ref{fig:5}
we present the probabilities
$P(d>\ell | {\rm trap})$
and
$P(d>\ell | {\rm ret})$
that the path of an RW on a directed ER network will be of length larger than $\ell$, 
given that it terminated by trapping or by retracing,
respectively. The results are presented for $N=1000$ and
$c=3$, $5$ and $7$.
The analytical results (solid lines) are found to be in excellent agreement
with the numerical simulations (symbols).
In both cases, the paths tend to become longer as $c$ is increased.
However, for each value of $c$, the paths which terminate due to retracing
are typically longer than the paths which terminate due to trapping.

Given that the path of an RW has terminated after $\ell$ steps, it is interesting
to evaluate the conditional probabilities 
$P({\rm trap} | d=\ell)$ 
and
$P({\rm ret} | d=\ell)$,
that the termination was caused by trapping or by
retracing, respectively.
Using Bayes' theorem,
these probabilities can be expressed by

\begin{equation}
P({\rm trap} | d=\ell) = \frac{p_{\rm trap} P(d=\ell | {\rm trap})}{P(d=\ell)},
\end{equation}

\noindent
and

\begin{equation}
P({\rm ret} | d=\ell) = \frac{p_{\rm ret} P(d=\ell | {\rm ret})}{P(d=\ell)}.
\end{equation}

\noindent
Clearly, these distributions satisfy
$P({\rm trap} | d=\ell) + P({\rm ret} | d=\ell) =1$.
Inserting the conditional probabilities
$P(d=\ell | {\rm trap})$
and
$P(d=\ell | {\rm ret})$
from Eqs.
(\ref{eq:p_lbeq})
and
(\ref{eq:p_lreq}),
respectively, we find that

\begin{equation}
P({\rm trap} | d=\ell) = 
\frac{P(d>\ell-1)}{e^{c} P(d=\ell)},
\label{eq:b_ell}
\end{equation}

\noindent
and

\begin{equation}
P({\rm ret} | d=\ell) = \left( 1-e^{-c} \right)
\frac{ \left[ c - c(\ell) \right]  }{c}
\frac{P(d>\ell-1)}{P(d=\ell)}.
\label{eq:b_ell2}
\end{equation}

\noindent
The corresponding distributions can be expressed in the form

\begin{equation}
P({\rm trap} | d>\ell) = 
\frac{\sum\limits_{t=\ell+1}^{N-1} 
P(d>t-1)}{ e^{c} P(d>\ell)},
\label{eq:b_ell3}
\end{equation}

\noindent
and

\begin{equation}
P({\rm ret} | d>\ell) = 
\left( \frac{1-e^{-c}}{c} \right) 
\sum_{t=\ell+1}^{N-1} 
\left[ c - c(t) \right] \frac{P(d>t-1)}{P(d>\ell)}.
\label{eq:r_ell}
\end{equation}

\begin{figure}
\centerline{
\includegraphics[width=16cm]{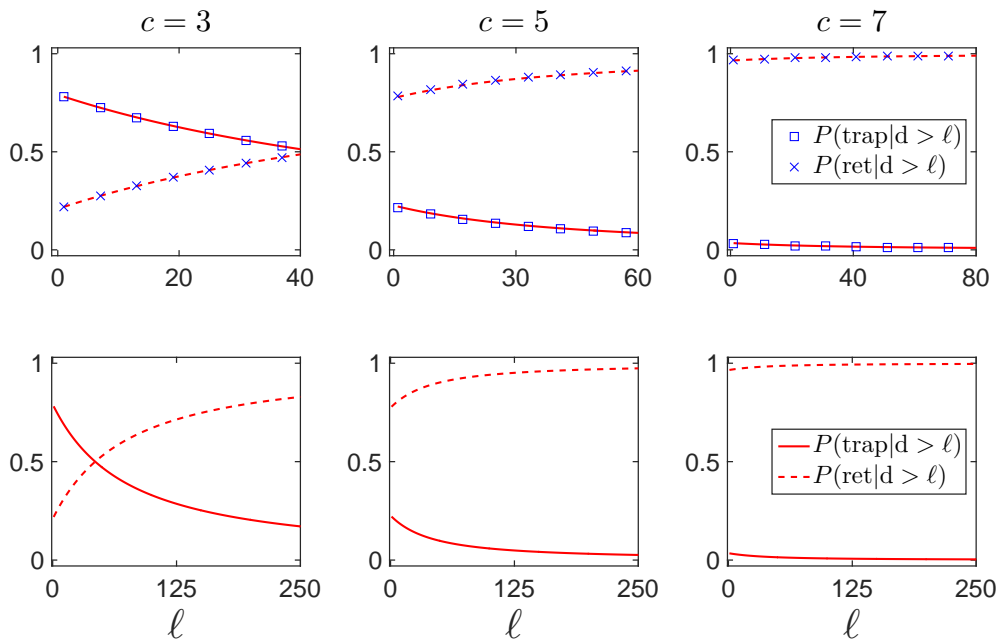}
}
\caption{
The conditional probabilities,
$P({\rm trap} | d>\ell)$
and
$P({\rm ret} | d>\ell)$,
that an RW path on a directed ER network 
will terminate by trapping or by retracing,
respectively, given that its length is larger than $\ell$,
are presented as a function of $\ell$.
Results are shown for an ER networks of size $N=1000$ and 
$c=3$, $5$ and $7$.
The theoretical results for
$P({\rm trap} | d>\ell)$ (solid lines)
are obtained from Eq.
(\ref{eq:b_ell3})
while the theoretical results for
$P({\rm ret} | d>\ell)$ (dashed lines)
are obtained from Eq.
(\ref{eq:r_ell}).
In the top row these results are compared to the results of numerical
simulations (symbols) finding excellent agreement. This comparison
is done for the range of path lengths which actually appear in the
numerical simulations and for which good statistics can be obtained.
Longer RW paths which extend beyond this range become extremely
rare, so it is difficult to obtain sufficient numerical data.
However, in the bottom row we show the theoretical results
for a larger range of path lengths. 
It is found that 
$P({\rm trap} | d>\ell)$
is a monotonically decreasing function of $\ell$
while
$P({\rm ret} | d>\ell)$ is monotonically increasing.
}
\label{fig:6}
\end{figure}

\noindent
These distributions also satisfy
$P({\rm trap} | d>\ell) + P({\rm ret} | d>\ell) =1$.
In Fig. 
\ref{fig:6}
we present the probabilities
$P({\rm trap} | d>\ell)$
and
$P({\rm ret} | d>\ell)$
that an RW path on a directed ER network will terminate due to trapping or retracing,
respectively, given that its length is larger than $\ell$.
Results are shown for directed ER networks of size $N=1000$ and 
$c=3$, $5$ and $7$.
The theoretical results for
$P({\rm trap} | d>\ell)$ (solid lines)
are obtained from Eq.
(\ref{eq:b_ell3})
while the theoretical results for
$P({\rm ret} | d>\ell)$ (dashed lines)
are obtained from Eq.
(\ref{eq:r_ell}).
As expected,
it is found that 
$P({\rm trap} | d>\ell)$
is a monotonically decreasing function of $\ell$
while
$P({\rm ret} | d>\ell)$ is monotonically increasing.
In the top row these results are compared to the results of numerical
simulations (symbols) finding excellent agreement. 
This comparison
is done for the range of path lengths which actually appear in the
numerical simulations.
Longer RW paths which extend beyond this range become extremely
rare, so it is difficult to obtain sufficient numerical data.
However, in the bottom row we show the theoretical results
for a larger range of path lengths. 
As can be seen,
for small values of $c$,
the curves of 
$P({\rm trap}|d>\ell)$
and
$P({\rm ret}|d>\ell)$
cross each other,
while for larger values of $c$
there is no such crossing.
In fact, long paths can be sampled using the pruned 
enriched Rosenbluth method, which was successfully used in the 
context of SAWs in polymer physics 
\cite{Grassberger1997}. 
In this method one samples long non-overlapping paths, 
keeping track of their weights, to obtain an unbiased sampling 
in the ensemble of all paths.

\section{The distribution of first hitting times on directed semi-ER networks}

In most of the directed networks encountered in nature the distribution of
in-degrees differs from the distribution of out-degrees. It is thus important
to disentangle the effects of the in-degrees and out-degrees on the distribution
of first hitting times. To this end, we extend out studies to a 
class of directed networks, referred to as directed semi-ER networks
in which the in-degrees follow a Poisson distribution, whose
mean is equal to $c$, as is the case in directed ER networks. However, the
out-degrees may be distributed according to any desired distribution with
the same mean, $c$. It is important to note that in the models 
studied here there are no correlations between
the in-degree and the out-degree of any given node.
Also, there are no degree-degree correlations between adjacent nodes.
They thus belong to the class of directed configuration model networks
\cite{Newman2001,Molloy1995,Newman2010,Fronczak2004,Annibale2009,Roberts2011}.

To construct an instance of a network from a given ensemble
of directed configuration model networks of $N$ nodes, one draws
the in-degrees and out-degrees from the desired 
distributions 
$p^{\rm in}(k)$ and $p^{\rm out}(k)$,
$k=0,1,\dots,N-1$,
producing the degree sequences
$k_i^{\rm in}$, 
and
$k_i^{\rm out}$,
$i=1,\dots,N$.
While the distributions must satisfy
the condition
$\sum_k k p^{\rm in}(k) = \sum_k k p^{\rm out}(k)$,
for each instance one needs to make sure that
$\sum_i k_i^{\rm in} = \sum_i k_i^{\rm out}$.
One then prepares the $N$ nodes such that each node, $i$, is 
connected to $k_i^{\rm in}$ incoming half links
and $k_i^{\rm out}$ outgoing half links
\cite{Newman2010}.
Pairs of an incoming half link from one node and an outgoing half
link from another node
are then chosen randomly
and are connected to each other in order
to form the network. 
The result is a network with the desired degree sequence and
no correlations.
Note that towards the end of the construction
the process may get stuck.
This may happen in case that the only remaining pairs of half links
are in the same node or in nodes which are already connected to each other.
In such cases one may perform some random reconnections 
in order to enable completion of the construction.

The directed semi-ER networks can be constructed using a simpler procedure
than the general configuration model networks.
Consider the adjacency matrix, $A$, of a directed semi-ER network of
size $N$. The matrix element $a_{ij}=1$ if there is a directed link
from node $i$ to node $j$, and zero otherwise.
The diagonal elements, $a_{ii}=0$, for all $i=1,2,\dots,N$.
The first step in the construction is to draw the sequence of out-degrees,
$k_i^{\rm out}$, $i=1,2,\dots,N$
from the distribution
$p^{\rm out}(k)$.
The $i$th row of $A$ thus includes $k_i^{\rm out}$ 1's and $N-k_i^{\rm out}$ 0's.
In the second step, one places randomly the $k_i^{\rm out}$ $1$'s
among the $N-1$ non-diagonal
matrix elements of the $i$th row. Since there are no correlations between the placements
of 1's in different rows, the in-degrees follow a Poisson distribution whose
mean is equal to $c$. In this procedure there is no need for any adjustments since 
the total number of incoming links is directly determined by the number of outgoing links.

Below we consider several examples of directed semi-ER networks.
In case that the out-degrees also follow the
Poisson distribution, the directed ER network is recovered. 
For integer values
of $c$, the out-degrees may follow a degenerate distribution, in
which the out-degrees 
$k_i^{\rm out}=c$ 
for all the nodes in
the network. Such networks may be considered as a hybridization
of an ER network and a regular graph
\cite{Bollobas2001}. 
Clearly, in these networks there are no
dead-end nodes of degree $k_i^{\rm out}=0$.

Another interesting example is the case in which the out-degrees
follow a power-law distribution of the form

\begin{equation}
p^{\rm out}(k) = A k^{-\gamma}, 
\label{eq:power_law}
\end{equation}

\noindent
for 
$k_{\rm min} \le k \le k_{\rm max}$,
where the lower cutoff
$k_{min} \ge 1$
and an upper cutoff 
$k_{max} \le N-1$.
Such distributions are also known as scale-free distributions.
The normalization coefficient, $A$ is given by

\begin{equation}
A = \frac{1}{\zeta(\gamma,k_{min})-\zeta(\gamma,k_{max}+1)},
\label{eq:power_lawcoeff}
\end{equation}

\noindent
where
$\zeta(s,a)$ is
the Hurwitz zeta function
\cite{Olver2010}.
The mean of the out-degree distribution,
$\langle k \rangle^{\rm out} = \sum_k k p^{\rm out}(k)$,
is given by

\begin{equation}
\langle k \rangle^{\rm out} =
\frac{\zeta(\gamma-1,k_{min})-\zeta(\gamma-1,k_{max}+1)}
{\zeta(\gamma,k_{min})-\zeta(\gamma,k_{max}+1)}.
\end{equation}

\noindent
%
%
In case that
$k_{\rm min} = 1$
the Hurwitz zeta function
$\zeta(s,1)$,
coincides with the Riemann zeta function, $\zeta(s)$
\cite{Olver2010}.
Therefore, in this case the mean of the out-degree distribution 
is given by

\begin{equation}
\langle k \rangle^{\rm out} = 
\frac{\zeta(\gamma-1)-\zeta(\gamma-1,k_{max}+1)}
{\zeta(\gamma)-\zeta(\gamma,k_{max}+1)}.
\label{eq:zetaR}
\end{equation}

\noindent
Since $\langle k \rangle^{\rm out}=c$, one can use
Eq. (\ref{eq:zetaR}) in order to obtain the value of the
exponent $\gamma$ which would yield the desired mean out-degree, $c$.
Such a directed semi-ER
network 
with a Poisson in-degree distribution and a power-law
out-degree distribution
can be considered as a hybridization of an ER 
network and a scale-free network
\cite{Barabasi2002}.
Since 
$k_{\rm min} = 1$,
these networks do not include any dead end nodes of degree
$k_i^{\rm out}=0$.

We also consider the case in which the out-degrees follow an exponential
distribution of the form

\begin{equation}
p^{\rm out}(k) = B e^{- \alpha k},
\end{equation}

\noindent
in the range
$k_{\rm min} \le k \le k_{\rm max}$,
where the lower cutoff is
$k_{\rm min} = 0$
and the upper cutoff is
$k_{\rm max} \le N-1$.
The normalization factor, $B$, is given by

\begin{equation}
B = \frac{1 - e^{-\alpha}}{1 - e^{-\alpha(k_{\rm max}+1)}}.
\end{equation}

\noindent
Due to the fast decay of the exponential distribution,
we can take the approximation in which
$k_{\rm max} \rightarrow \infty$.
In this case the normalization factor is simplified to
$B = 1 - e^{-\alpha}$.
Within this approximation,
the mean of the exponential out-degree distribution is given by

\begin{equation}
\langle k \rangle^{\rm out} = \frac{1}{e^{\alpha} - 1}.
\end{equation}

\noindent
Since $\langle k \rangle^{\rm out}=c$,
we find that in order to obtain a desired value of the
mean out-degree, $c$, the exponent $\alpha$ of the exponential
out-degree distribution should be given by

\begin{equation}
\alpha = \ln \left( \frac{c+1}{c} \right).
\end{equation}

\noindent
The fraction of nodes in these networks which are dead-end nodes
of zero out-degree is given by $p^{\rm out}(k=0) = B$,
which in the limit of large $k_{\rm max}$ is well approximated by

\begin{equation}
p^{\rm out}(k=0) =1-e^{-\alpha}.
\label{eq:pout}
\end{equation}

In order to calculate the distribution of first hitting times of RWs on directed
semi-ER networks we recall that the probability that an RW will visit
node $i$, which resides on the subnetwork of the yet-unvisited nodes,
is proportional to its in-degree, $k_i^{\rm in}(t)$.
The expectation value of the number of incoming links of the node
visited by the RW at time $t$, which originate from the subnetwork of the yet-unvisited
nodes, are removed at that time is $c(t)$.
Since the distribution of out-degrees is uncorrelated with the distribution
of in-degrees, the expectation value of the number of outgoing links 
from the node visited at time $t$ to the 
subnetwork the yet-unvisited nodes, which are then removed,
is also given by $c(t)$.
Therefore, the analysis presented in Sec. 4,
which is based on special properties of the Poisson distribution,
still holds for the in-degree distribution,
which remains Poisson, and its mean, $c(t)$ is given by Eq.
(\ref{eq:coft}).

The probability of not terminating by retracing at time $t$,
$P_{\rm ret}(d>t | d>t-1)$,
given by Eq.
(\ref{eq:P_r2}),
is equal to the ratio between the mean degree 
$c(t)$ of the subnetwork of the yet-unvisited nodes and the
mean degree of the entire network, $c$.
Since the time evolution of $c(t)$ is not affected by the
out-degree distribution, the probability of termination
by retracing remains identical to the case of an ER network
with the same value of $c$.

For RWs on a directed ER network,
the probability of not terminating by trapping,
$P_{\rm trap}(d>t | d>t-1)$,
given by Eq.
(\ref{eq:trap2}),
is independent of the time, $t$.
For RWs on directed semi-ER networks, 
it is replaced by

\begin{equation}
P_{\rm trap}(d>t | d>t-1) = 1 - p^{\rm out}(k=0),
\end{equation}

\noindent
where 
$p^{\rm out}(k=0)$
is the probability that a randomly chosen node in the
network is a dead-end node which does not have any outgoing links.
Since this probability does not evolve in time,
it can be determined from the initial distribution
$p^{\rm out}(k)$.
As shown above, for the degenerate distribution
and for the power-law distribution
the probability
$p^{\rm out}(k=0) =0$.
Therefore, the trapping scenario does not apply to RWs on 
these networks and their paths terminate only by retracing.
On the other hand, 
for the exponential distribution
the probability
$p^{\rm out}(k=0)$
is non-zero and is given by Eq.
(\ref{eq:pout}).
Thus, for the network with an exponential degree
distribution, 
$P_{\rm trap}(d>\ell) = \exp(- b \ell)$,
where
$b = - \ln [1 - p^{\rm out}(k=0)]$.

\begin{figure}
\centerline{
\includegraphics[width=18cm]{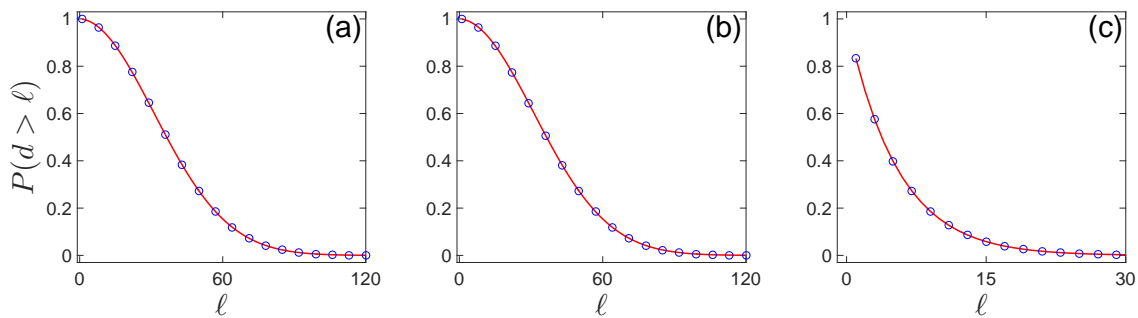}
}
\caption{
The tail distributions 
of first hitting times, 
$P(d > \ell)$, 
of RWs on directed semi-ER networks
of size $N=1000$ and $c=5$,
in which the in-degrees follow the Poisson distribution
and the out-degrees are distributed according to a
degenerate (regular) distribution (a), a power-law
(scale-free) distribution with
$\gamma=2.14$ and $k_{\rm min}=1$ (b),
and an exponential distribution with 
$\alpha=0.18$ and $k_{\rm min}=0$ (c).
The analytical results
(solid lines)
are in excellent agreement with 
the results obtained
from numerical simulations (circles).
}-
\label{fig:7}
\end{figure}

In Fig.
\ref{fig:7} 
we present the
tail distributions 
of first hitting times, 
$P(d > \ell)$, 
of RWs on three directed 
semi-ER networks, all of them
of size $N=1000$ and $c=5$.
In these networks
the in-degrees follow a Poisson distribution, 
while the out-degrees are distributed according to a
degenerate (regular) distribution (a), 
a power-law 
(scale-free)
distribution,
which is obtained using the parameters
$\gamma=2.14$ 
and 
$k_{\rm min}=1$ 
(b),
and an exponential distribution,
obtained using the parameters
$\alpha=0.18$
and
$k_{\rm min}=0$ 
(c).
The analytical results
(solid lines)
are in excellent agreement with 
the results obtained
from numerical simulations (circles).

\begin{figure}
\centerline{
\includegraphics[width=8cm]{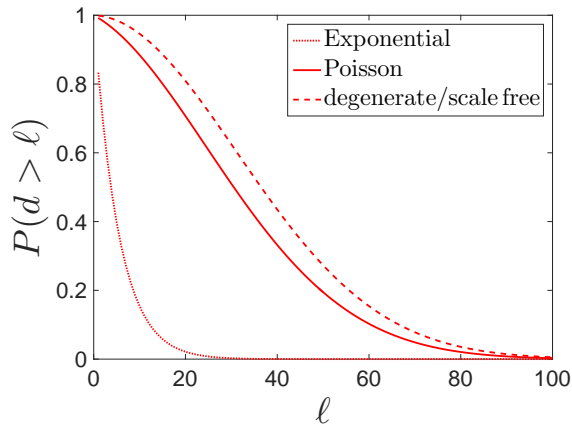}
}
\caption{
Analytical results for 
the tail distributions of first hitting times,
$P(d>\ell)$, 
of RWs on
four directed networks of size 
$N=1000$ and $c=5$: 
a directed ER network (solid line),
in which both the in-degrees and the out-degrees follow 
a Poisson distribution,
and three directed semi-ER networks,
in which the
in-degrees follow a Poisson distribution 
and the out-degrees are
distributed according to 
an exponential distribution (dotted line),
with $\alpha=0.18$ and $k_{\rm min}=0$,
a degenerate (regular) distribution (dashed line),
and a power-law (scale-free) distribution (dashed line),
with $\gamma=2.14$ and $k_{\rm min}=1$.
The results obtained for the two latter distributions are 
identical, and thus shown by the same dashed line.
}
\label{fig:8}
\end{figure}

In Fig.
\ref{fig:8} 
we present
analytical results for 
the tail distributions of first hitting times,
$P(d>\ell)$, 
of RWs on 
directed 
semi-ER networks 
of size $N=1000$ and $c=5$,
in which the
out-degrees are
distributed according to a
Poisson distribution (solid line),
an exponential distribution (dotted line)
as well as 
a degenerate distribution
(regular graph) and a power-law distribution
(which coincide with each other and shown by a dashed line).
The results for the degenerate distribution and the power-law
distribution are identical.
This is due to the fact that in both networks the termination
scenario by trapping does not exist, while the termination
probabilities by retracing are identical in the two networks.
This result is surprising in light of the fact that
the degenerate distribution and the power-law distribution
are entirely different from each other.
In particular, the degenerate distribution is infinitely narrow,
while the power-law distribution exhibits a broad tail and
for $\gamma \le 3$ its variance diverges.

The RW paths on networks with Poisson and exponential out-degree 
distributions may terminate either by retracing or by trapping. The
network with an exponential out-degree distribution exhibits a much 
larger fraction of dead-end nodes. Therefore the first hitting
times of RWs on this network are much shorter than on the
network with a Poisson out-degree distribution 
(which is, in fact, an undirected ER network).

\section{Summary and discussion}

We presented analytical results for   
the distribution of first hitting times of 
RWs on directed
ER networks.
Starting from a random initial node,
these RWs hop randomly along directed edges
between adjacent nodes 
until their paths terminate.
Termination may occur either by retracing or by trapping.
In the retracing scenario the RW
steps into a node which it has already visited
before.
In the trapping scenario 
the RW becomes trapped in a 'dead-end' node which has
no outgoing edges. 
The number of steps pursued from the initial node
up to the termination of the RW path is called 
the first hitting time.
Using recursion equations
we obtained analytical results for the 
tail distribution of first hitting times, 
$P(d>\ell)$ as well as for the mean, median and 
standard deviation of this distribution.
The results are found to be in excellent agreement with numerical
simulations.
It was found that the tail distribution
$P(d > \ell)$
can be expressed as a product of an exponential distribution 
and a Rayleigh distribution.

\begin{figure}
\centerline{
\includegraphics[width=8cm]{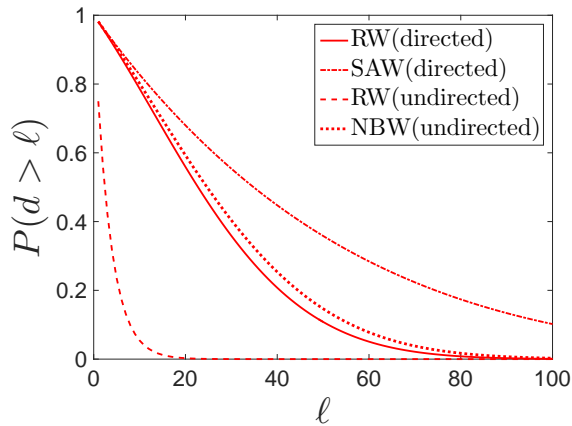}
}
\caption{
Analytical results for 
the tail distribution of first hitting times,
$P(d>\ell)$, 
of RWs on a
directed ER network (solid line),
obtained from Eq.
(\ref{eq:tail_noapp}).
For comparison we also present the tail distribution of first hitting times
of RWs
on an undirected ER network (dashed line),
obtained using the analytical expression derived in
Ref. 
\cite{Tishby2016b},
as well as of NBWs on undirected
ER networks (dotted line),
obtained using the analytical expression derived in
Ref. 
\cite{Tishby2016c}.
In all three cases the network size is
$N=1000$ and $c=4$.
It is found that the first hitting times of RWs on directed ER
networks are much longer than those of RWs on undirected networks.
This is due to the fact that in undirected ER the backtracking scenario
is a dominant termination mechanism, while in directed ER network 
this scenario does not exist and is replaced by the trapping scenario
which is much weaker.
The first hitting times of NBWs on undirected ER 
networks are found to be slightly longer than those of RWs on
directed ER networks.
The distribution of last hitting times of SAWs on a directed
ER network,
obtained from Eq.
(\ref{eq:P(d>l)}),
is also shown (dashed-dotted line). 
As expected, the last hitting times are significantly longer than the
first hitting times on the same network.
}
\label{fig:9}
\end{figure}

In Fig. 
\ref{fig:9}
we present analytical results
(solid line)
for the tail distribution of first hitting times,
$P(d > \ell)$,
of RWs on a
directed ER network of size
$N=1000$ and $c=4$.
For comparison we also present the
corresponding distribution of RWs 
on an undirected ER network (dashed line)
with the same value of $c$
(based on Ref. \cite{Tishby2016b}).
It is found that the first hitting times of RWs on directed ER
networks are much longer than those of RWs on undirected ER networks.
This is due to the fact that in undirected networks the backtracking scenario
is a dominant termination mechanism which contributes
to a significant reduction in the first hitting times.

For the sake of completeness,
we briefly summarize the properties of
first hitting times of RWs on undirected
ER networks.
The conditional probability associated with the retracing scenario
of RWs on undirected ER networks is given by
 
\begin{equation}
P_{\rm ret}(d>t | d>t-1)
=\frac{c(t)+1}{c+1}.
\end{equation}

\noindent
This probability is somewhat larger than the result for directed ER networks,
given by Eq.
(\ref{eq:P_r2}).
However, the main
difference between the distributions of first hitting times
in the directed and undirected ER networks is the fact that
on directed networks the backtracking scenario does not exist.
The conditional probability associated with the backtracking scenario
for RWs on undirected ER networks is given by
\cite{Tishby2016b}

\begin{equation}
P_{\rm backtrack}(d>t | d>t-1) = 
\frac{c-1+e^{-c}}{c}.
\end{equation}

\noindent
Thus, backtracking is an important termination scenario, particularly
for sparse networks.
Unlike the case of undirected networks,
in directed ER networks the backtracking scenario does not exist 
because the backwards link to the previous node exists only 
with probability $p$, namely it is as likely as a link to any other node.
However, another termination scenario emerges. 
This is the trapping scenario, which occurs when the RW enters a dead-end node
which does not have any outgoing link and becomes trapped in that node. 
It turns out that the trapping scenario in directed ER networks
is much less likely to happen 
than the backtracking scenario in the corresponding undirected ER network.
The conditional probability associated with the trapping scenario for RWs
on directed ER networks is given by

\begin{equation} 
P_{\rm trap}(d>t | d>t-1) = 1-e^{-c}.
\end{equation}

\noindent
Note that both the backtracking and trapping probabilities do not
depend on time.
However, their dependencies on the mean degree, $c$ is very different
from each other.
While the backtracking 
probability essentially decreases as $1/c$,
the trapping probability decreases exponentially with $c$. 
Therefore, the trapping mechanism is 
much less likely to occur and so the paths of RWs on directed
ER networks are much longer than on a corresponding undirected
networks with the same value of $c$. 

The properties of RWs on directed ER networks resemble those of 
non-backtracking RWs (NBWs) on undirected ER networks
\cite{Tishby2016c}.
This is due to the fact that in both cases  the termination of the
RW path by the backtracking scenario is suppressed.
In Fig. 
\ref{fig:9}
we also show the tail distribution
of NBWs on undirected
ER networks (dotted line).
Indeed, it is found to be very similar to the tail distribution of RWs on
directed ER networks.
However, the tail distribution of the NBWs is shifted
slightly to the right compared to the tail distribution of the corresponding
RW on a directed ER network.
This is due to the fact that the probability of retracing is slightly lower
for the NBW compared to the RW on a directed ER network.
More precisely,
for an NBW on an undirected ER network the probability
of retracing at the $t+1$ time step is given by
$[c-c(t)]/(c+1)$, 
compared to 
$[c-c(t)]/c$ for an RW on a directed ER networks.
For completeness, we also show in Fig.
\ref{fig:9}
the distribution of last hitting times of SAWs on directed ER networks
(dashed-dotted lines),
obtained from Eq.
(\ref{eq:tail2}).
As expected, the last hitting times are much longer than the first hitting times.

We performed a detailed analysis of the probabilities,
$p_{\rm trap}$ and $p_{\rm ret}$,
that the termination will take place via the trapping or via
the retracing mechanism, respectively.
We obtained analytical expressions for these probabilities in terms
of the network size, $N$, and the mean degree, $c$.
We also obtained analytical expressions for the conditional 
distributions of the path lengths,
$P(d=\ell | {\rm trap})$ 
and
$P(d=\ell | {\rm ret})$
for the paths which terminate by
trapping or by retracing,
respectively.
Finally, we obtained analytical expressions for 
the conditional probabilities
$P({\rm trap} | d=\ell)$
and
$P({\rm ret} | d=\ell)$
that a path which terminates after $\ell$ steps
is terminated by trapping or by retracing, respectively.
It was found that the two termination mechanisms exhibit
very different behavior.
Since the initial node is chosen to be a node with at
least one outgoing link, 
the trapping probability sets in starting from the 
second step. The trapping probability is constant throughout the path.
As a result, the trapping mechanism alone would produce a 
geometric distribution of path lengths.
The retracing mechanism also sets in starting from the 
second step and its rate increases linearly in time.
The balance between the two termination mechanisms 
depends on the mean degree
of the network.
In the limit of sparse networks, 
the trapping mechanism is dominant and most
paths are terminated long before the retracing mechanism becomes relevant.
In the case of dense networks, the trapping probability is low and most
paths terminate by the retracing mechanism.
These results provide useful insight into the general problem of 
survival analysis and the statistics of mortality 
or failure rates,
under conditions in which two or more 
failure mechanisms coexist
\cite{Finkelstein2008,Gavrilov2001}.

We have shown that the approach developed in this paper
applies not only to ER networks but also to directed semi-ER networks
in which the in-degree distribution is a Poisson distribution
while the out-degrees may follow any desired distribution
which has the same mean, $c$, as the in-degree distribution.
To demonstrate this result we presented the tail distribution,
$P(d>\ell)$, for directed semi-ER networks with degenerate (regular), 
power-law (scale free)
and exponential out-degree distributions.
It was shown that the rate of termination by retracing is
determined by the Poisson in-degree distribution, which 
controls the temporal evolution of $c(t)$. The rate of termination
by trapping is determined by the out-degree distribution,
or more specifically by the  
probability $p^{\rm out}(k=0)$ that a randomly chosen node
will have no outgoing link.

In a broader context, the distributions of first hitting times
and last hitting times of RWs are examples of a broad class of
distributions of path lengths and first passage times in random 
networks
\cite{Redner2001}.
A related distribution, which provides useful information on
the underlying structure of the network, is the distribution of shortest
path lengths between random pairs of nodes
\cite{Newman2001,Blondel2007,Dorogotsev2003,Hofstad2007,Hofstad2008,Esker2008,Katzav2015,Nitzan2016,Melnik2016}.
Examples of such distributions involving RWs are the distribution
of first passage times between random pairs of nodes
\cite{Sood2005}
and
the distribution of cover times
\cite{Kahn1989}.
We expect the methodologies developed in this paper to be
useful for the study of other structural and dynamical distributions
in random networks.

\newpage


\section{Appendix A: Detailed calculation of $P_{\rm ret}(d>\ell)$}

In this Appendix we present the detailed evaluation of
$P_{\rm ret}(d>\ell)$.
Taking the logarithm of 
$P_{\rm ret}(d>\ell)$,
as expressed in Eq.
(\ref{eq:cond6}),
we obtain

\begin{equation}
\ln \left[P_{\rm ret}(d>\ell)\right] = 
\sum_{t=1}^{\ell} \ln \left[ \frac{c(t)}{c} \right].
\label{eq:Pret_sum}
\end{equation}

\noindent
Replacing the sum by an integral we obtain

\begin{equation}
\ln \left[ P_{\rm ret}(d>\ell) \right]
=
\int_{1/2}^{\ell+1/2} 
\ln \left[\frac{c\left(t\right)}{c}\right]dt.
\label{eq:Pret_int}
\end{equation}

\noindent
Plugging in the expression for $c(t)$ 
from Eq.
(\ref{eq:coft})
and rearranging terms in the integrand
we obtain

\begin{equation}
\ln \left[ P_{\rm ret}\left(d>\ell\right) \right]
=\int_{1/2}^{\ell+1/2}
\ln\left[1-\frac{t}{\left(N-1\right)}\right]dt.
\end{equation}

%

\noindent
After integration,
replacement of $N-1$ by $N$
and rearrangement of terms we obtain 

\begin{equation}
P_{\rm ret}\left(d>\ell\right)
\simeq
\exp\left[\left( \ell+\frac{3}{2}-N \right)
\ln\left(1-\frac{\ell+1/2}{N-1}\right)-\ell-\frac{1}{2}\right].
\end{equation}

\noindent
In approximating the sum of 
Eq. (\ref{eq:Pret_sum}) by the integral of Eq. (\ref{eq:Pret_int})
we have used the formulation of the middle Riemann sum. 
Since the function
$\ln[P_{\rm ret}(d>\ell)]$ is a monotonically decreasing function, the value of
the integral is over-estimated by the left Riemann sum, $L(\ell)$, and under-estimated
by the right Riemann sum, $R(\ell)$. 
The error involved in this approximation is thus 
bounded by the difference $\Delta(\ell) = L(\ell) - R(\ell)$,
which satisfies
$\Delta(\ell) = -\ln( 1 - \ell/N) $.
Thus, the relative error in $P_{\rm ret}(d>\ell)$ due to the approximation of the sum
by an integral is bounded by
$\eta_{\rm SI} \sim \ell/N$.
Comparing the values obtained from the sum and the integral
over a broad range of parameters, we find that the pre-factor of
the error is very small, so in practice the error introduced by approximation of the
sum by an integral is negligible.


\newpage


\begin{thebibliography}{10}

\bibitem{Lawler2010}
Lawler G F and Limic V 2010
{\it Random Walk: A Modern Introduction}
(Cambridge: Cambridge University Press)

\bibitem{Rudnick2010}
Rudnick J and Gaspari G 2010
{\it Elements of the Random Walk: An introduction for Advanced Students and Researchers}
(Cambridge: Cambridge University Press)

\bibitem{Kampen2007}
van Kampen N G 2007
{\it Stochastic Processes in Physics and Chemistry}
(Amsterdam: North Holland)

\bibitem{Redner2001}
Redner S 2001 
{\it A Guide to First Passage Processes}
(Cambridge: Cambridge University Press)


\bibitem{Montroll1965}
Montroll E E and Weiss G H 1965
{J. Math. Phys.} {\bf 6} 167




\bibitem{Debacco2015}
De Bacco C, Majumdar S N and Sollich P 2015 
{\it J. Phys. A} {\bf 48} 205004




\bibitem{Sood2005}
Sood V, Redner S and ben-Avraham D 2005
{\it J. Phys. A} {\bf 38} 109

\bibitem{Kahn1989}
Kahn J D, Linial N, Nisan N and Saks M E 1989 
{\it J. Theor. Probab.} {\bf 2} 121  


\bibitem{Herrero2003}
Herrero C P and Saboy\'a M 2003
{\it Phys. Rev. E} {\bf 68} 026106


\bibitem{Herrero2005}
Herrero C P 2005
{\it Phys. Rev. E} {\bf 71} 016103 

\bibitem{Herrero2005b}
Herrero C P 2005
{\it J. Phys. A} {\bf 38} 4349



\bibitem{Herrero2007}
Herrero C P 2007
{\it Eur. Phys. J. B.} {\bf 56} 71


\bibitem{Tishby2016b}
Tishby I, Biham O and Katzav E 2016
{\it J. Phys. A} {\bf 50} 115001

\bibitem{Erdos1959}
Erd{\H o}s P and R\'enyi 1959 
{\it Publ. Math.} {\bf 6}  290 

\bibitem{Erdos1960}
Erd{\H o}s P and R\'enyi 1960
{\it Publ. Math. Inst. Hung. Acad. Sci.} 
{\bf 5} 17

\bibitem{Erdos1961}
Erd{\H o}s P and R\'enyi 1961
{\it Bull. Inst. Int. Stat.} 
{\bf 38} 343

\bibitem{Madras1996}
Madras N and Slade G 1996
{\it The Self Avoiding Walk}
(Boston: Birkh\"auser)

\bibitem{Slade2011}
Slade G 2011
{\it Surveys in Stochastic Processes}, 
Proceedings of the 33rd SPA Conference in Berlin, 2009, 
EMS Series of Congress Reports,
eds. Blath J, Imkeller P, and Roelly S

\bibitem{Tishby2016a}
Tishby I, Biham O and Katzav E 2016
{\it J. Phys. A} {\bf 49} 285002


\bibitem{Bollobas2001}
Bollobas B 2001
{\it Random Graphs, Second Edition}
(London: Academic Press)



\bibitem{Bang-Jensen2007}
Bang-Jensen J and Gutin G 2007
{\it Digraphs Theory, Algorithms and Applications}
(Berlin: Springer-Verlag)


\bibitem{Graham2008}
Graham A J and Pike D A 2008
{\it Atlantic Electronic Journal of Mathematics}
{\bf 3}, 1


\bibitem{Havlin2010}
Havlin S and Cohen R 2010
{\it Complex Networks: Structure, Robustness and Function}
(Cambridge University Press, New York).



\bibitem{Dorogovtsev2001}
Dorogovtsev S N, Mendes J F F and Samukhin A N 2001
{\it Phys. Rev. E} {\bf 64} 025101

\bibitem{Newman2001}
Newman M E J, Strogatz  S H and Watts D J 2001
{\it Phys. Rev. E} {\bf 64} 026118





\bibitem{Olver2010} 
Olver F W J, Lozier D W, 
Boisvert R F and Clark C W 2010
{\it NIST Handbook of Mathematical Functions} 
(Cambridge: Cambridge University Press) 




\bibitem{Gompertz1825}
Gompertz B 1825
{\it Philosophical Trans. R. Soc. London A} {\bf 115} 513


\bibitem{Johnson1995}
Johnson N L, Kotz S and Balakrishnan N 1995
{\it Continuous Univariate Distributions}
(New York: John Wiley \& Sons)



\bibitem{Shklovskii2005}
Shklovskii B I 2005
{\it Theory in Biosciences} {\bf 123} 431

\bibitem{Ohishi2008}
Ohishi K, Okamura H and Dohi T 2009
{\it Journal of Systems and Software} {\bf 82} 535




\bibitem{Pitman1993}
Pitman J 1993
{\it Probability} 
(New York: Springer-Verlag)

\bibitem{Grassberger1997}
Grassberger P 1997
{\it Phys. Rev. E} {\bf 56} 3682 


\bibitem{Molloy1995}
Molloy M and Reed B 1995
{\it Random Struct. Algorithms} {\bf 6} 161 

\bibitem{Newman2010}
Newman M E J 2010 
{\it Networks: an Introduction} 
(Oxford: Oxford University Press).

\bibitem{Fronczak2004}
Fronczak A, Fronczak P and Holyst J A 2004 
{\it Phys. Rev. E} {\bf 70} 056110 

\bibitem{Annibale2009}
Annibale A, Coolen A C C, Fernandes L P, Fraternali F and Kleinjung J 2009
{\it J. Phys. A} {\bf 42} 485001

\bibitem{Roberts2011}
Roberts E S, Schlitt T and Coolen A C C 2011
{\it J. Phys. A} {\bf 44} 275002

\bibitem{Barabasi2002}
Albert R and Barab\'{a}si A L 2002
{\it Rev. Mod. Phys.} {\bf 74} 47 

\bibitem{Tishby2016c}
Tishby I, Biham O and Katzav E 2016
arXiv:1609.08375, accepted to {\it J. Phys. A}

\bibitem{Finkelstein2008}
Finkelstein M 2008
{\it Failure Rate Modelling for Reliability and Risk}
(Springer-Verlag, London) 

\bibitem{Gavrilov2001}
Gavrilov L A and Gavrilova N S 2001
{\it J. theor. Biol} {\bf 213} 527

\bibitem{Blondel2007}
Blondel V D, Guillaume J -L, Hendrickx J M and Jungers R M 2007 
{\it Phys. Rev. E} {\bf 76} 066101 

\bibitem{Dorogotsev2003}
Dorogotsev S N, Mendes J F F and Samukhin A N 2003 
{\it Nuclear Physics B} {\bf 653} 307 

\bibitem{Hofstad2007}
van der Hofstad R, Hooghiemstra G and Znamenski D 2007 
{\it Electronic Journal of Probability} 
{\bf 12} 703 

\bibitem{Hofstad2008}
van der Hofstad R and  Hooghiemstra G 2008 
{\it J. Math. Phys.} {\bf 49} 125209 

\bibitem{Esker2008}
van der Esker H, van der Hofstad R and Hooghiemstra G 2008 
{\it J. Stat. Phys.} {\bf 133} 169 

\bibitem{Katzav2015}
Katzav E,  Nitzan M, ben-Avraham D, Krapivsky P L, K\"uhn R, 
Ross N and Biham O 2015
{\it EPL} {\bf 111} 26006 

\bibitem{Nitzan2016}
Nitzan M, Katzav E, K\"uhn R and Biham O 2016,
{\it Phys. Rev. E} {\bf 93} 062309




\bibitem{Melnik2016}
Melnik S and Gleeson J P 2016
arXiv:1604.05521


\end{thebibliography}
\end{document}